\documentclass[%
 reprint,
 amsmath,amssymb,
 aps,
]{revtex4-2}

\usepackage{graphicx}
\usepackage{dcolumn}
\usepackage{bm}
\usepackage[caption = false]{subfig} 

\begin{document}

\title{Beam dynamics performance of the proposed PETRA IV storage ring }

\author{I. Agapov}
 \email{ilya.agapov@desy.de}
\author{S.~A.~Antipov}
\author{R. Bartolini}
\author{R. Brinkmann}
\author{Y.-C. Chae}
\author{E.~C.~Cort\'es~Garc\'ia}
\author{D.~Einfeld}
\author{T. Hellert}
\author{M. H\"uning}
\author{M. A. Jebramcik}
\author{J. Keil}
\author{C. Li}
\author{L. Malina}
\author{R. Wanzenberg}
\affiliation{%
Deutsches Elektronen-Synchrotron DESY, Notkestr. 85, 22607 Hamburg, Germany 
}%

\date{\today}

\begin{abstract}
The PETRA IV project for upgrading the 2.3 km 6 GeV PETRA III storage ring to a diffraction-limited synchrotron radiation source is nearing the end of its detailed technical design phase. We present the ring lattice based on the hybrid six-bend achromat (H6BA) cell and a detailed evaluation of its beam dynamics performance. Design challenges as well as unique opportunities associated with a low emittance ring of a large size are discussed. 

\end{abstract}

\maketitle


\section{\label{sec:intro} Introduction}

The PETRA III synchrotron radiation facility \cite{Balewski:2004iz} has been in operation since 2009, being at the time of construction the 6 GeV source with the lowest emittance in the world. With the advent of synchrotron light sources based on the multi-bend achromat lattices (\cite{Einfeld:xe5006, Tanaka:2016vtg, Leemann:2009zz, LEEMANN201833, PhysRevAccelBeams.26.091601, Steier:2019sbc,Borland:2018jgq}), and especially with the successful implementation of the ESRF-EBS (\cite{Revol:2018eiq, PhysRevAccelBeams.24.110701,esrf2023}), an upgrade of the PETRA facility to stay on the frontline of x-ray science has been necessary \cite{Wanzenberg:2017imm, Schroer:ig5056}. Several lattice designs have been evaluated in the conceptual design phase (see \cite{Agapov:2018mhn, Keil:2018hyc, Keil:2021yag}), with the conceptual design published in 2019 based on on the hybrid seven-bend achromat lattice \cite{Schroer2019}. The large ring circumference makes PETRA IV the ring with the smallest achievable emittance among the existing or proposed fourth-generation light sources, but at the same time makes its beam dynamics aspects most challenging. In this paper, we present the lattice option on which the technical design of the facility is based \cite{Agapov:2022nqg, Bartolini:2022qfn}, beam dynamics aspects including nonlinear lattice optimization, tolerances and commissioning simulations,  evaluation of collective instabilities, and the achievable performance parameters.

The paper is organized as follows. In the rest of the introduction, design objectives and challenges are highlighted. The PETRA IV facility is briefly introduces in Section \ref{sec:facility}. Section \ref{sec:lattice} presents the lattice design following the H6BA concept \cite{Raimondi:2023rby}, and Section \ref{sec:dynamics}, central to this paper, presents a detailed study of single-particle and collective beam dynamics aspects and challenges. Section \ref{sec:parameters} lists parameters of main subsystems such as magnets, RF, and the injector complex. Finally, Section \ref{sec:advanced} discusses the possibility of exploiting the facility size and extremely low emittance for several advanced radiation generation techniques.

\subsection{Lattice design goals and constraints}

The next generation of photon science experiments would greatly benefit from hard x-ray (10-50 keV) photon beams with a high degree of transverse coherence and brightness levels in excess of $10^{22}$ phot./mm/mrad/0.1\%BW. 
These unprecedented levels of brightness and coherence are achieved by using improved undulator technologies such as cryogenic in-vacuum or superconducting devices (\cite{BAHRDT2018149}), but first of all, by generating electron beams of extremely low (tens of pm rad) emittance. These levels are achievable in a 6 GeV synchrotron of 2.3 km circumference such as PETRA by employing the multi-bend achromat lattices. While theoretically very low emittances (in the few pm rad range) are possible, the design parameters are set taking realistic constraints into account, that will be discussed further. Taking these constraints into account, the goals were set to deliver emittances of below 30 pm rad and beam currents of up to 200 mA.

\begin{figure}[b!]
\includegraphics[width=.95\linewidth]{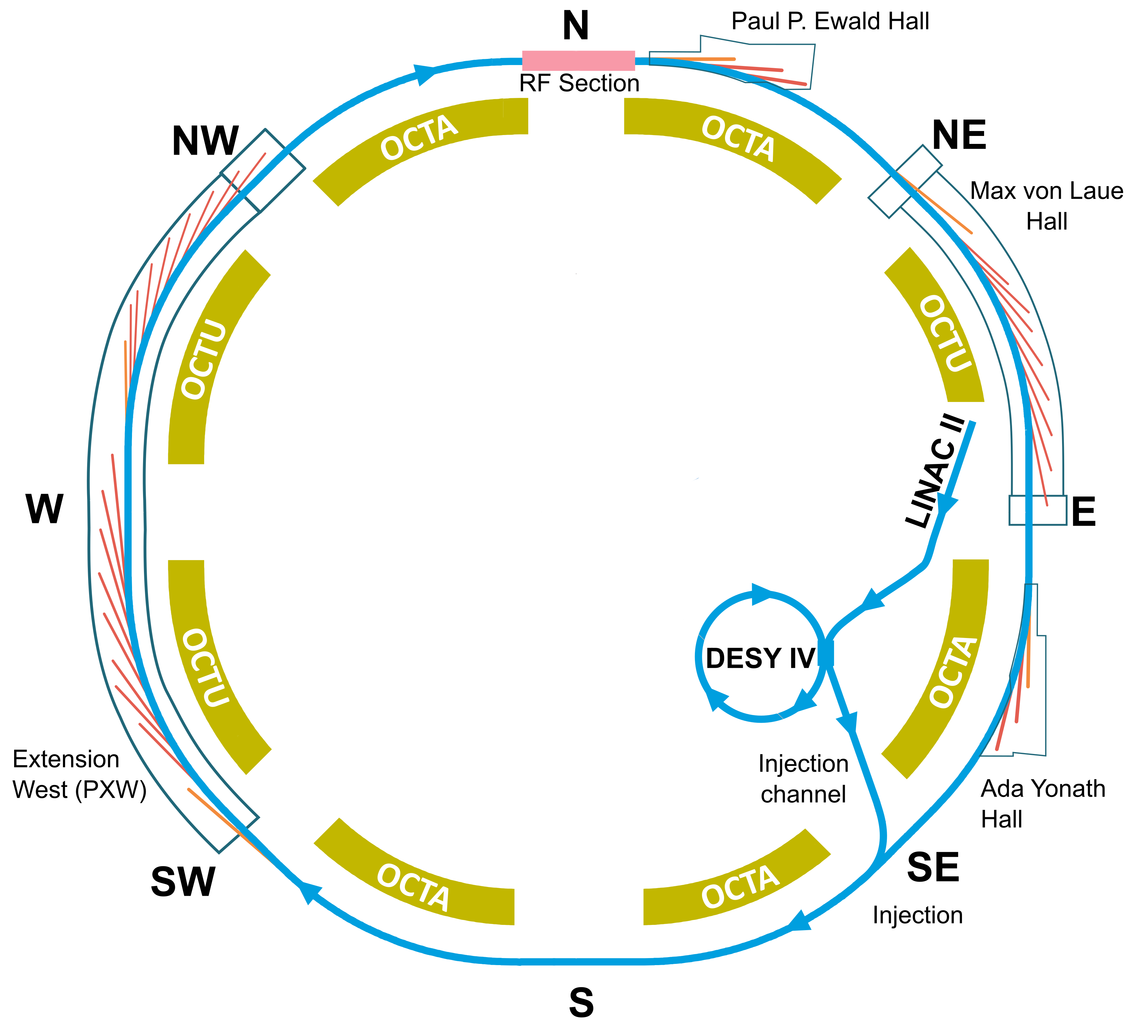}
\caption{\label{fig_layout} Layout of the PETRA IV facility. Existing experimental halls (Max von Laue, Peter P. Ewald, and Ada Yonath) will be reused. An additional experimental hall ("Extension West") will be constructed. }
\end{figure}

In PETRA IV, similar to PETRA III, only part of the lattice can be equipped with insertion devices. PETRA~IV will feature a new experimental hall with additional beam-lines significantly increasing the total space available for insertion devices (see Fig. \ref{fig_layout}). Outside of the experimental halls, i.e. for about half of the circumference, the machine should follow the existing tunnel. 
The tunnel has a width of only about 3.1 m, which together with the need for cables, escape routes, and other infrastructure elements, constrains the machine geometry transversely to an envelope of about 10 cm.

PETRA III was designed to make most use of the limited number of ID straights. This was achieved by, first, having a short DBA (\cite{dba1,dba2}) cell of ca. 23 m length that has  sextupoles removed, with the chromaticity correction distributed to the FODO cells of the rest of the ring; and second, by extensively exploiting the so-called canting, i.e. operating two insertion devices in one straight, with a corrector magnet used to introduce an angle (of 1 mrad, 5 mrad, or 20 mrad depending on location) in the electron trajectory between two devices, thus separating the radiation cones and allowing multiple beam-lines per insertion straight section.

While preserving the whole arrangement of source points is impossible when no significant emittance deterioration is allowed, many beam-lines can be kept, significantly simplifying the logistics when a ca. 23 m cell is adopted. For those several beam-lines that feature a 20 mrad canting angle, the dispersion generated in the straight is such that the influence of insertion devices on emittance is prohibitively strong. An estimate can be easily made using the radiation integrals \cite{Sands:1969lzn}, and it could be shown that the influence of canting on the emittance scales as $\propto \beta_x \theta_x^2$ where $\beta_x$ is the horizontal beta function in the middle of the ID straight, and $\theta_x$ the canting angle. Canting angles larger than ca. 5 mrad are not compatible with the low emittance ring design.

The constraint on the cell length noticeably complicates the machine optimization. In the absence of such a constraint, one possible path of nonlinear dynamics optimization would have been to increase the cell length at the cost of a somewhat larger emittance (see e.g. \cite{Wolski}). But this path has been ruled out based on the requirements on partial source point preservation and the total number of beam-lines.

Only moderate technological advances wrt.\ e.g.\ achievable magnet gradients are permissible to allow project implementation in the nearest future with minimal R\&D effort on magnet technology.

A new booster synchrotron is required. The existing infrastructure also severely constrains the booster geometry and transfer line layout. The booster design is outside of the scope of the paper, the injection scheme is briefly discussed in 
Section \ref{sec:injection_chain}.

\subsection{Key challenges}

The design objective of the PETRA IV lattice is to maximize the brightness delivered by a portfolio of insertion devices. As with all low-emittance ring designs, there is a number of trade-offs to be considered:

The most significant technical limitation in the low emittance ring design is the maximum achievable quadrupole and sextupole strength. Without this limit (and neglecting any nonlinear dynamics limitations to be discussed later) the natural emittance can be made almost arbitrarily small. The maximum gradient is limited by the field saturation limits of commonly available magnetic materials. The field gradients can be further increased by reducing the bore radius. With decreased bore radius the implementation of the vacuum system becomes challenging and the effect of impedance increases. These considerations lead to limiting the maximum achievable magnet strength to about 115 T/m and the minimum bore radius to about 9-10 mm. 

Another important factor is the relative length of insertion devices with respect to the ring circumference (filling factor). Since the emittance is generated in the arcs, its minimization could be achieved by reducing the filling factor, while for maximizing the experimental throughput larger filling factor would be beneficial. In practice, ID straight section length of approx. 5 m for a cell length of approx. 23--25 m is the best compromise for PETRA~IV.

Further, emittance can be minimized by either extensively exploiting damping wigglers, or creating lattices with a large partition shift (i.e. shifting the damping from the longitudinal to the transverse plane). Both approaches can at the same time lead to increased beam energy spread. The energy spread is detrimental to the brilliance. The exact effect depends on parameters of the insertion device, the experiment performed, and the x-ray optics, and is not discussed here. Energy spreads above 0.1\% are generally undesirable. 

Special attention should be paid to the assessment of errors on the machine performance.
All light sources based on MBA lattices suffer from increased sensitivity to alignment errors: strong focusing quadrupoles in conjunction with strong sextupoles to compensate the large natural chromaticity of these lattices create substantial feed-down effect and machine instability with alignment errors that are below what is realistically achievable. So-called machine bootstrapping is required to set up and run the machine. Demonstration of this procedure is necessary for all future projects, and the experience of MAX IV and ESRF-EBS showed that these procedures are adequate and the design parameters can be adjusted in relatively short time. Nevertheless, the error analysis played an important role in the PETRA IV lattice selection, as will be discussed later in Section \ref{sec:tolerance}.

\section{PETRA IV Facility Overview}
\label{sec:facility}

In this section the overall facility design is briefly summarized.
The facility consists of the photon science complex (beam-lines and experiments), the storage ring and the injector complex. The storage ring feeds up to approx. 30 undulator insertions (photon beam can be further split to allow more experimental stations). The storage ring will operate in two modes: brightness mode with 1920 stored bunches (4 ns spacing) with the total current of 200 mA and the timing mode with 80 bunches and total current of 80 mA. Other operation modes consistent with 2 ns minimum bunch spacing, single bunch current limitation of approx. 2 mA and total current limitation of 200 mA are conceivable.

Intra-beam scattering and Touschek effects contribute significantly to the emittance growth and the decrease of beam lifetime. These effects are mitigated by having sufficiently large number of buckets with a 500 MHz RF system and single bunch lengthening with a 3rd harmonic (1.5 GHz) system. 

The injector complex features a new booster, a refurbished S-band 450 MeV linac, and a 450 MeV accumulator ring. As an alternative for possible future upgrades, plasma injector \cite{PhysRevAccelBeams.24.111301, Antipov:2022ifd,Pousa:2021sua} R\&D is being pursued to potentially reduce the injector complex footprint and the energy consumption.

Injection into the storage ring is done in the standard off-axis accumulation mode, with fast strip-line kickers that can minimize the perturbation of the experiments. 

Among other notable design features are the fully NEG-coated vacuum chambers, fast orbit and transverse multi-bunch feedback systems, and extensive use of permanent magnet dipoles.

\section{The H6BA Lattice}
\label{sec:lattice}

The storage ring has a geometry inherited from the HEP programme of PETRA in the 1970s, which is unusual for a synchrotron radiation facility. It has eight arcs, four straight sections of approx. 108~m length, and four straight sections of approx. 64~m length. 

Each arc is composed of nine hybrid six-bend achromat (H6BA, \cite{Raimondi:2023rby}) cells (see Figures \ref{fig_cell}, \ref{fig_cell_cad}) each. Moreover, some of the long straight sections feature insertion devices of approx.~10~m length. Special triplet optics is used to focus the beam in this insertion (see Figure \ref{fig_lss}).

\subsubsection*{Achromats}

Two cell types are used for the eight octants. One cell type features a user insertion device (ID), while the other type has a damping wiggler (DW) insertion. The bending angle of each achromat is 5$^{\circ}$, the total number of achromats is~72. Due to geometrical reasons the cell length of these two cell types are slightly different.

To keep some of the positions of source points of existing undulator beamlines of PETRA~III in the Max von Laue Hall, a cell length of 23~m has to be used there. This will avoid costs for relocating existing beamlines. Achromats with a length of 23~m will also be used in the new Extension Hall West (PXW).

In five octants a shorter cell length of 22.75~m is required due to geometrical constraints of the existing tunnel. In these cells damping wigglers will be installed. These cells have an identical magnet arrangement with a shorter ID straight. In addition the strength of quadrupole magnets up- and downstream of the DWs have to be changed to make the phase advances and the beta functions of both cell types nearly equal.

\begin{figure}[t]
\includegraphics[width=.95\linewidth]{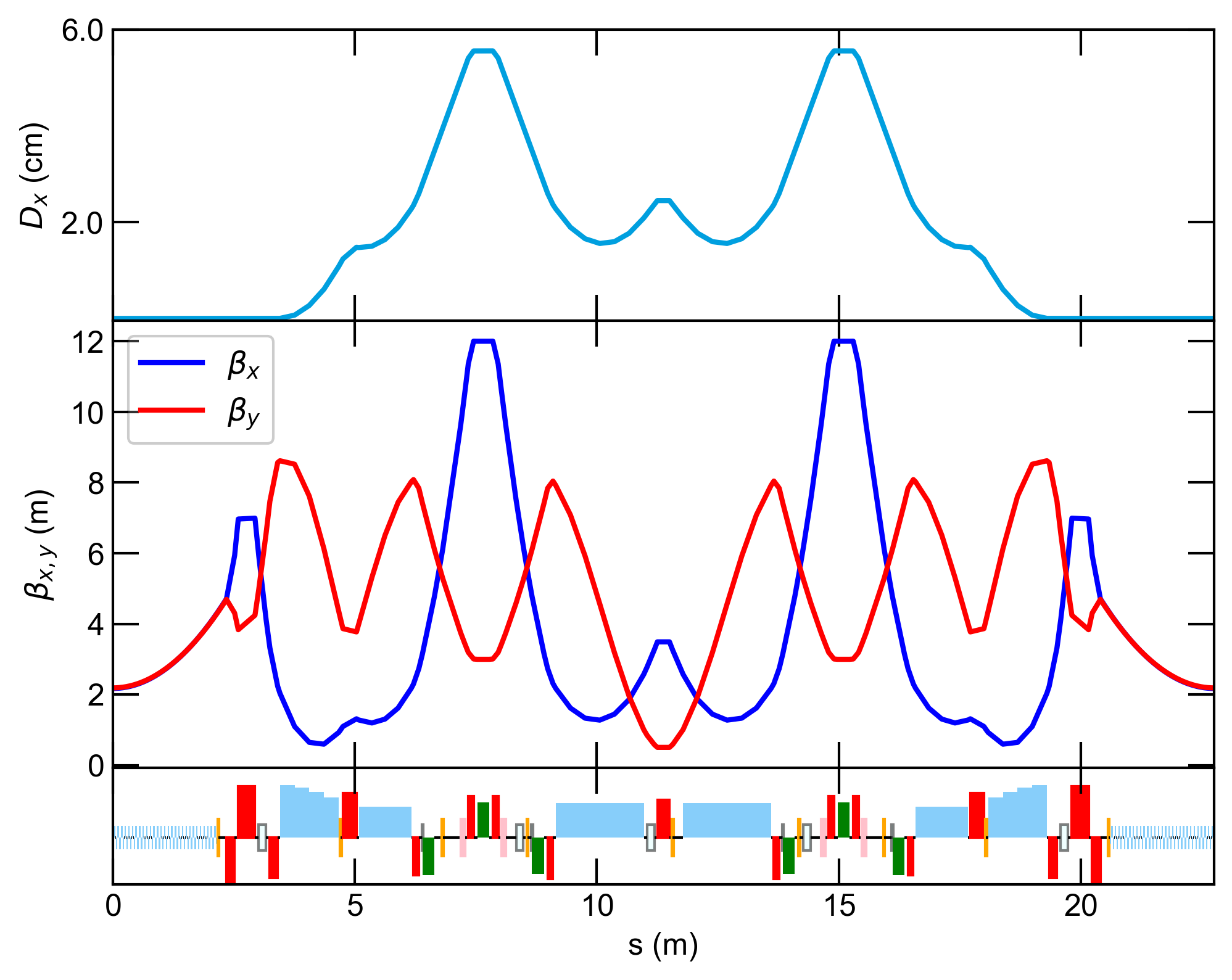}
\caption{\label{fig_cell} Layout of the H6BA cell. Quadrupoles are shown in red, dipoles in blue, sextupoles in green, octupoles in pink, BPMs in orange, and orbit correctors as transparent boxes. The right- and leftmost elements are the two halves of an insertion device. }
\end{figure}

\begin{figure}[t]
\includegraphics[width=.95\linewidth]{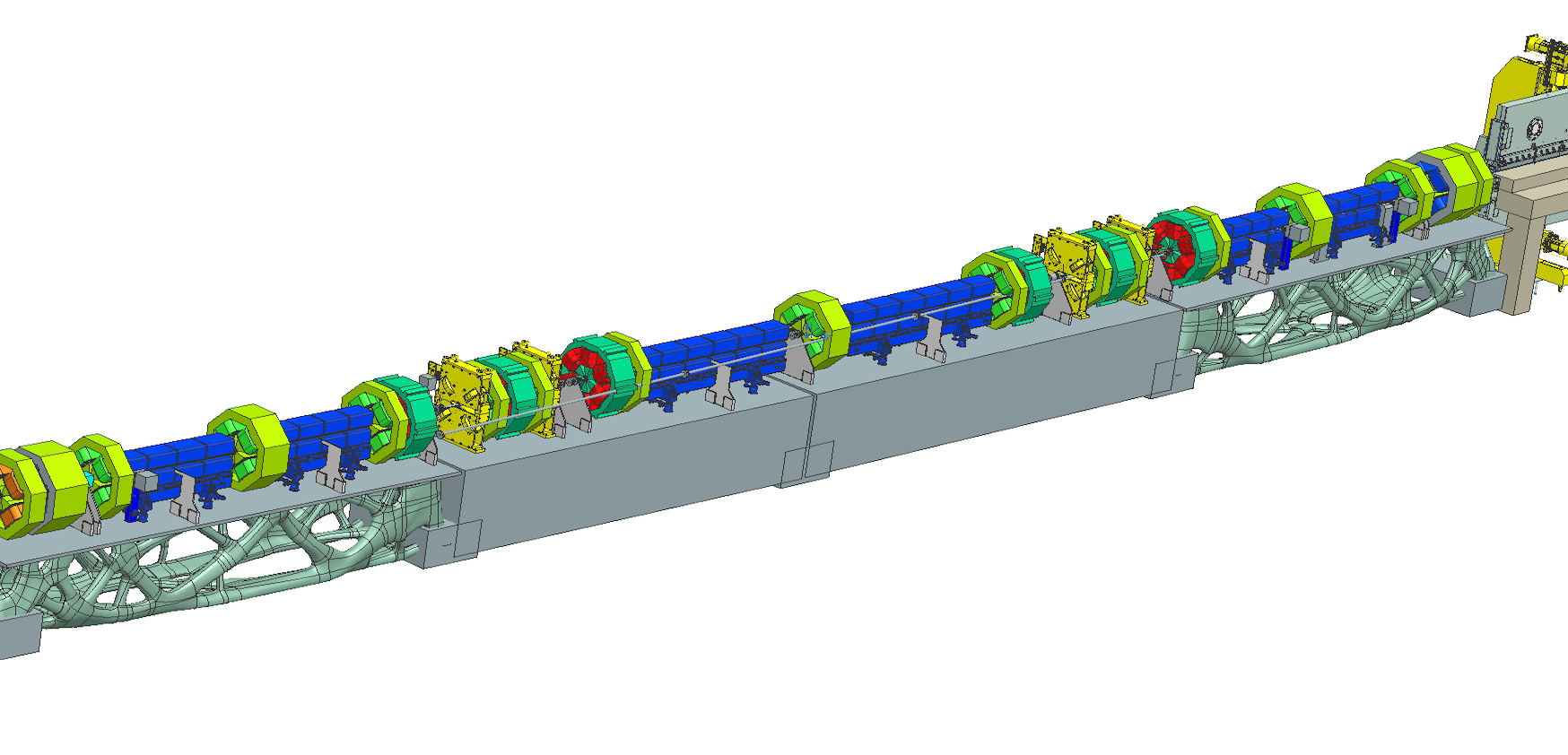}
\caption{\label{fig_cell_cad} CAD view of the H6BA cell. Bending magnets are shown in blue, quadrupoles in light green, sextupoles in dark green, and octupoles in yellow.}
\end{figure}

The optics and the arrangement of elements of the H6BA cell is shown in Fig.~\ref{fig_cell} and a CAD view of the cell in Fig.~\ref{fig_cell_cad}. 

A quadrupole triplet up- and downstream of the ID straight is used to focus the beta functions to $\beta_x = \beta_y = 2.2$~m in the center of the ID which is close to the optimum value yielding maximum brightness. The dispersion function at the IDs is zero to avoid emittance contribution of the undulators when the gaps are closed. As a compromise between small beta functions and a feasible quadrupole design the maximum gradient in the triplet was limited to 115~T/m. The other quadrupoles in the achromat have gradients of 100~T/m or less. 

Between the insertion straight and the section for chromaticity correction there are two dipoles with both longitudinal and transverse gradients (combined-function magnets). A focusing quadrupole is in between. All dipole magnets have a vertical defocusing field. This  makes the cell  more compact and helps to increase the horizontal damping partition number $J_x$ and reduce the emittance. Both dipoles consists of four permanent magnet blocks.

The chromaticity correction section has a dispersion bump which consists of a symmetric arrangement of four quadrupoles with three sextupoles and two octupoles in between. The chromatic sextupoles are placed near the peaks of horizontal and vertical beta functions. The octupoles are installed near large horizontal beta function and dispersion function to correct mainly the second order chromaticity.

Between the dispersion bump and a focusing quadrupole in the center of the cell there is another combined-function dipole which is longer compared to the other two dipoles. It consists of six blocks of permanent magnets.

The cell is reflected mirror-symmetrically. The betatron phase advance between the groups of sextupoles is close to $\pi$ in both planes. There are nine beam position monitors, and seven orbit correctors per cell per plane.

\subsubsection*{Damping Wigglers}

The damping wigglers are considered to be part of the cell optics, and bring down the lattice emittance from approx.~43~pm~rad to 20~pm~rad. This allows to have less aggressive optics compared to e.g. seven-bend achromat lattices: ultra-large circumference of the PETRA ring makes the peak dispersion function of a seven-bend achromat lattice smaller, thus resulting in the need for stronger sextupoles and inferior beam dynamics. The damping wigglers allow to recover the emittance at the cost of RF power and some energy spread growth.  
During operation of PETRA~IV the undulators of the users will contribute partly to the reduction of the emittance if their gaps are closed. For the case that all gaps of undulators are open around 40~damping wigglers would be needed to achieve 20 pm~rad. These damping wigglers have a length of 4.44~m, a peak field of a sin-like field of 0.85~T and a period length of 74~mm. Some of them will have variable gaps to be used for an emittance feedback to compensate gap movements of user IDs when the gaps are changed during operation.
The number of variable-gap wigglers is limited by the available installation space, currently about five of such devices are foreseen, with the rest of the wigglers having fixed gaps. None of the wigglers can be employed as a dedicated radiation source due to the lack of space for beam-line installation.
Note that the energy loss per turn is dominated by either the damping wigglers or the user insertion devices (see Table \ref{tab:lattice}), and the previously considered seven-bend achromat lattice of PETRA IV   featured similar requirements on the RF voltage (8 MV, \cite{Schroer2019}), while reaching a horizontal emittance of 8 pm rad at the low bunch current limit assuming the full set of insertion devices.
Beam dynamics considerations, in particular the improvement in momentum acceptance from approx. 1 \% \cite{Keil:2021yag} to approx 3 \% (as discussed later), prevailed over a more aggressive emittance optimization in the lattice selection.

\subsubsection*{Straight Sections}

\begin{figure}
\includegraphics[width=.95\linewidth]{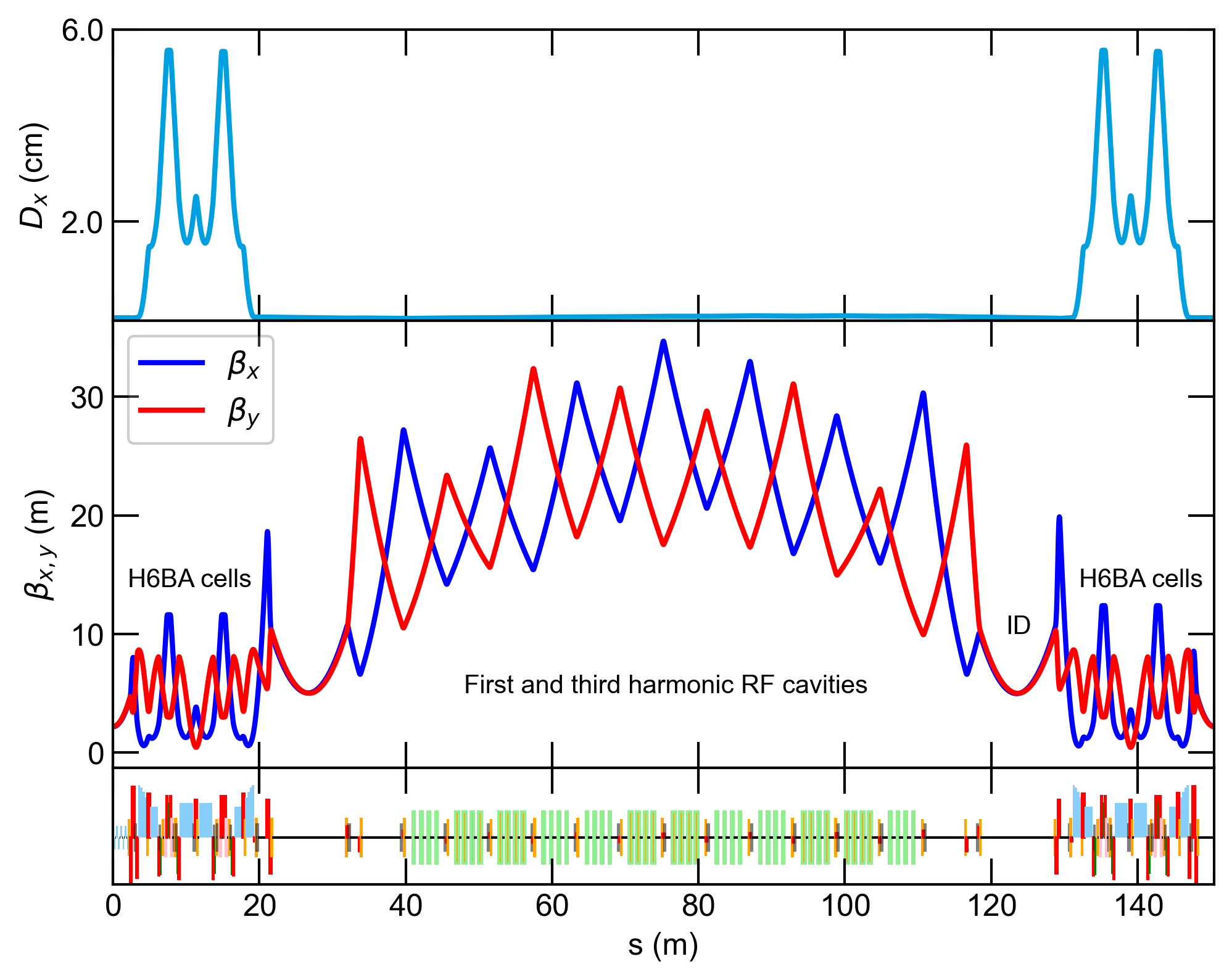}
\caption{\label{fig_lss} A long straight section featuring 10 m long low-beta insertions for the flagship IDs (North straight).  }
\end{figure}

\begin{figure}
\includegraphics[width=.95\linewidth]{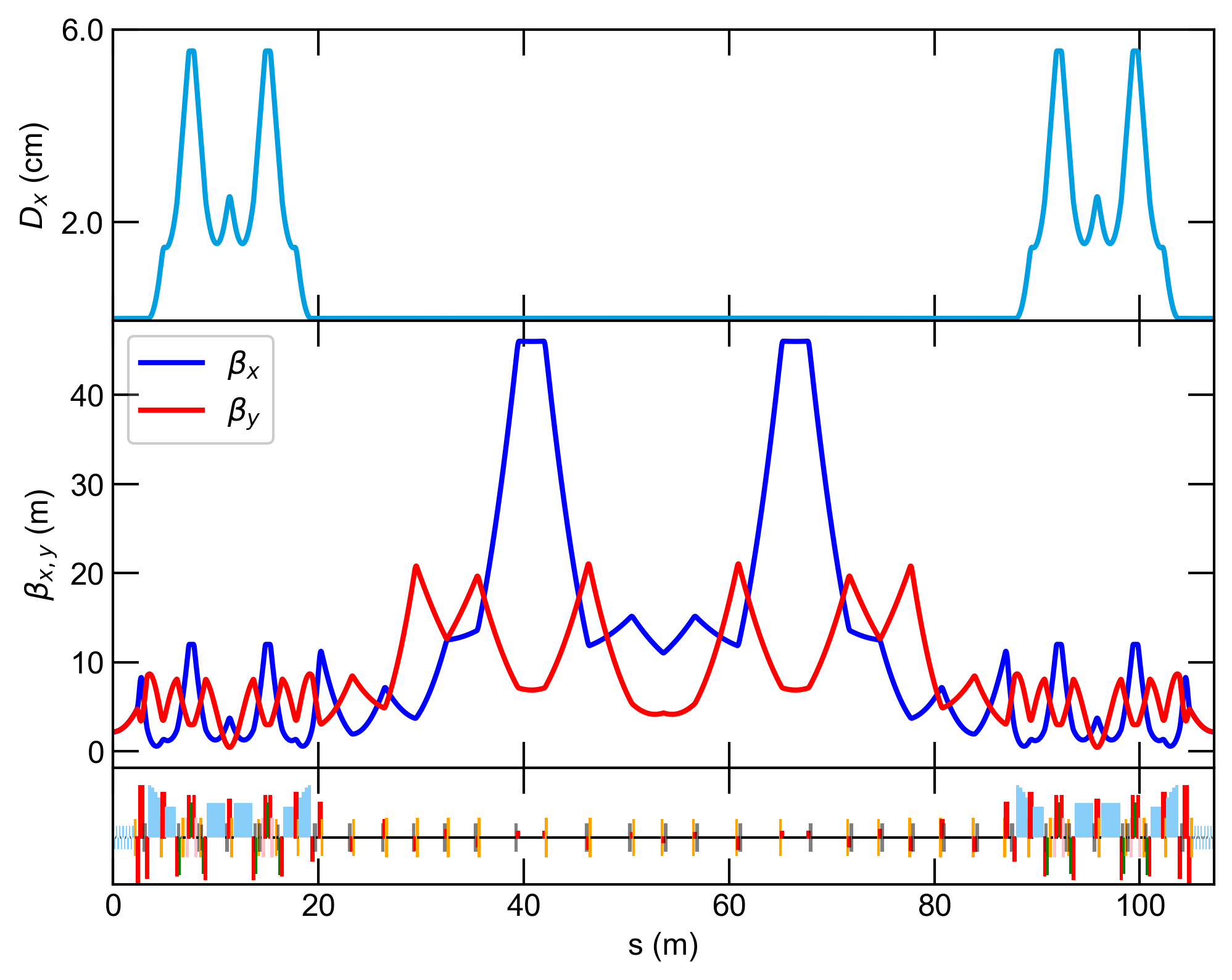}
\caption{\label{fig_injs} Injection straight (South-East).}
\end{figure}

The eight arcs are connected by the long straight sections of different types. These straight sections are matched to minimize their impact on the beam dynamics, as discussed further. The long straights come in several types. 
First, there are three long straight sections (N, W, E) and three shorter long straight sections (SW, NW, NE) that comprise low-beta insertions at the beginning and the end of the straight, and FODO-like matching in between. The long and short version of such straights have similar design, and the long version is shown in Figure~\ref{fig_lss}. The injection section in South-East is shown in Figure~\ref{fig_injs}. It features a peak of the horizontal beta function of 46~m where the injection septum is placed, thus minimizing the footprint of the septum blade on the acceptance. The RF will occupy the straight section North. The straight section South has a simple FODO structure and will be used for collimation. 

For the straight sections with low-beta insertions, 10 m space would be available for an insertion device. 
The design of the straight section is symmetric to facilitate the beam dynamics optimization, but only the downstream section can be used to accommodate an insertion device, since the photon beam from the upstream insertion would diverge too much by the time it would reach a potential extraction location.  
As the photon and electron beams are optimally matched when the beta function in the middle of the insertion device is approx. $\beta^{*} \approx L_{ID}/\pi$ where $L_{ID}$ is the undulator length \cite{Walker:PRAB.22.050704}, a $\beta^{*} $ of 4 m is not significantly different from the optimal value of 3.2 m for a 10 m insertion device. Reducing the $\beta^{*}$ below 4 m results in difficulties with preserving the optimal phase advance in the long straight section required for the nonlinear beam dynamics optimization. Moreover, as discussed in Section \ref{sec:advanced}, there is a trade-off between the undulator length and the minimal achievable gap, so a somewhat shorter device could be more beneficial for high-brightness application, while a 10 m device could better suit high-flux applications. Investigations of these issues are presently ongoing in the framework of the definition of the PETRA IV beamline portfolio.

\begin{figure*}
\includegraphics[width=.99\linewidth]{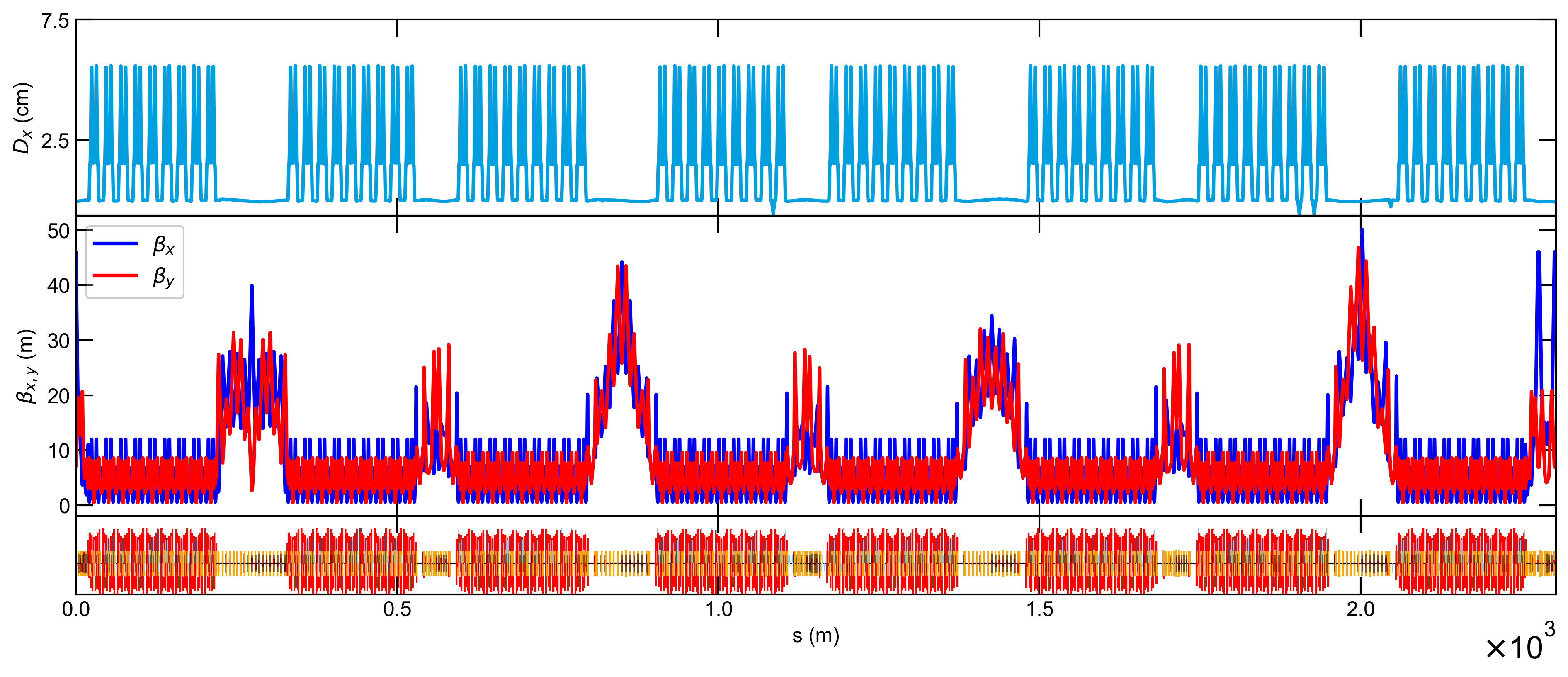}
\caption{\label{fig_ring_latti} Optical functions of the H6BA lattice. Injection point corresponds to the longitudinal coordinate $s=0$.}
\end{figure*}

\begin{table}
\caption{\label{tab:lattice} Parameters of the H6BA lattice. The parameters with damping wigglers (DW) correspond to an estimated configuration of insertion devices. }
\begin{ruledtabular}
\begin{tabular}{ll}
Parameter & Value  \\
\hline
Tunes $Q_x$,$Q_y$ & 164.18, 68.27  \\
Natural chromaticity $\xi_x$,$\xi_y$ & -230, -196  \\
Corrected chromaticity $\xi_x$,$\xi_y$ & 6, 6  \\
Momentum compaction factor $\alpha_C$ & $3.3 \times 10^{-5}$  \\
Standard ID space & 4.9 m  \\
$\beta_{x,y} $ at ID, standard cell  & 2.2 m, 2.2 m  \\
$\beta_{x,y} $ at ID, flagship IDs  & 4 m, 4 m  \\
Nat. hor. emittance $\varepsilon_x$ no DW, zero current & 43 pm rad \\
Nat. hor. emittance $\varepsilon_x$ with DW, zero current & 20 pm rad \\
Rel. energy spread $\delta_E$ no DW, zero current	& $0.7 \times 10^{-3}$  \\
Rel. energy spread $\delta_E$ with DW, zero current	& $0.9 \times 10^{-3}$  \\
Energy loss per turn, lattice no DW	& 1.3 MeV  \\
Energy loss per turn, lattice with DW	& 4.0 MeV  \\
\end{tabular}
\end{ruledtabular}
\end{table}

\subsubsection*{Canted ID straights}

The horizontal beta function of 2.2~m at the center of the ID straight of the H6BA cell is substantially smaller compared to previous lattice designs for PETRA~IV based on H7BA-achromats (\cite{Keil:2018hyc, Keil:2021yag}). This is an advantage if canted ID straights are used. It allows larger canting angles $\theta_x$ of canted ID straights (or more canted ID straights with small angles), as the emittance contribution of two undulators in a canted ID straights scales with $\theta_x^2 \beta_x$. Currently, using several 5 m ID straight sections with canting angles of $\theta_x = \pm 2.5$ mrad and one flagship ID straight section (West) with a canting angle $\theta_x = \pm 0.5$ mrad is foreseen.

The optical functions of the combined ring are shown in Figure~\ref{fig_ring_latti}. The lattice parameters are presented in Table~\ref{tab:lattice}. The damping wiggler period and field strength are such that their impact on emittance and energy spread is comparable to a typical user insertion device.  The parameters in Table~\ref{tab:lattice} correspond to an estimated average configuration of insertion devices. The trade-offs between installing the full set of wigglers to provide the target beam parameters in the initial operation stages with a reduced set of user insertion devices and installing a reduced set of wigglers initially are under discussion with the experimental community.

\section{Beam dynamics}
\label{sec:dynamics}

\subsection{Analytical lattice optimization}

Achieving an emittance of several tens of pm rad by using a multi-bend achromat (MBA) design is relatively easy for PETRA~IV. The large circumference allows to use many bending magnets with small deflection angles. However, the strong focusing of quadrupoles required in the MBA cell produces a large natural chromaticity. In addition, the small dispersion function of only a few cm reduces the effectiveness of the sextupoles. 

Strong sextupoles are necessary to correct the natural chromaticity to a slightly positive value. They will reduce the dynamic aperture (DA) and the momentum acceptance (MA). As a consequence the injection efficiency is low and the Touschek lifetime short. Because of weaker sextupoles the hybrid MBA design with two dispersion bumps a modification of which has been adopted by all of the latest 6 GeV synchrotron radiation sources  ~\cite{Revol:2018eiq, Raimondi:2023rby, Borland:2018jgq,Jiao:2018kke} has a clear advantage for PETRA~IV compared to the higher order achromat designs that are more suitable for smaller rings with smaller beam energy compared to PETRA~IV and more stringent requirements on momentum acceptance due to stronger Touschek scattering.

Besides the objective of minimum emittance, the non-linear properties of the lattice have to be optimized to achieve sufficient DA and MA. Different measures to reduce the impact of the strong sextupoles on the nonlinear dynamics have been included in the design. As linear optics and non-linear dynamics are coupled in the strong focusing MBA lattices, a compromise has to be found between emittance optimization and non-linear dynamics performance.

\subsubsection*{On-Momentum Optimization}
 

A phase advance of $\pi$ in both planes between a sextupoles pair with equal Twiss functions and a linear map in between compensates all phase dependent geometric resonance driving terms (RDTs) of first order in sextupole strength. Since $\alpha_{x,y}=0$ at the center of the dispersion bump and a symmetric arrangement of sextupoles the phase advances between corresponding sextupole pairs in the two dispersion bumps are almost equal. However, the interleaved sextupole arrangement and the finite length of the sextupoles breaks the exact cancellation. 

The phase advances per cell for the two different cell types were chosen to be equal. Changes of the phase advances and optical functions due to the opening and closing of the gaps of IDs and damping wigglers are corrected  locally by quadrupoles close to the ID. Quadrupoles between the sextupole pairs are not used for that purpose to preserve the $\pi$-phase advance condition. 

To create a higher super-periodicity of PETRA~IV, the phase advances of the long and short straight sections are set to $2\pi$ in both planes so that the full ring has a super-periodicity of 72 for on-momentum electrons. However, for off-momentum electrons the super-symmetry is not fulfilled anymore as the chromaticity contributions of the straights and the achromats are different. 


\subsubsection* {Off-Momentum Optimization}

To prevent that the lifetime of PETRA~IV is dominated by losses due to the Touschek scattering, a large momentum acceptance (MA) of the lattice is necessary. Besides a large off-momentum DA this implies that the momentum dependent change of the optical functions $D_x(s,\delta)$, $D'_x(s,\delta)$, $\beta_{x,y}(s,\delta)$ and $\alpha_{x,y}(s,\delta)$ are as small as possible, where $\delta = \Delta{}p/p$ is the relative momentum deviation. Touschek-scattered particles will have large momentum offsets of several percent and start with an initial horizontal coordinate vector relative to the reference orbit of $(x,x') = (x_0 + D_x(s)\delta + \partial{}D_x(s)/\partial\delta \times\delta^2, x'_0 + D_x'(s)\delta + \partial{}D_x'(s)/\partial\delta \times\delta^2)$ to the second order in $\delta$. In addition, the off-momentum particles have distorted optical functions of $\beta_{x,y}(s,\delta) = \beta_{x,y} + \partial{}\beta_{x,y}(x)/\partial\delta\times\delta$ and similar for $\alpha_{x,y}(s,\delta)$. 

Compared to a seven-bend achromat cell of similar length, the H6BA cell has already smaller chromatic beta functions $\partial{}\beta_{x,y}(x)/\partial\delta$ and chromatic dispersion function $\partial{}D_x(x)/\partial\delta$ because of weaker sextupoles. This helps to reduce the contributions to the second order chromaticity $\xi_{x,y}^{(2)}$ of quadrupoles and sextupoles. In the dispersion bump four octupoles are installed at locations with $\beta_x > \beta_y$ and large $D_x$ to reduce mostly the horizontal 2nd order chromaticity $\xi_x^{(2)}$ as their contribution scales with $b_4D_x^2\beta_{x}$ \cite{Leemann:PhysRevSTAB.14.030701}.

The short and long straight sections between the eight arcs should not increase the optimized off-momentum optical functions of the achromat. For this reason the Montague chromatic amplitude function $W_{x,y}(s)$ \cite{Montague:443342} and dispersion functions $\partial{}D_x(x)/\partial\delta$ of the straights were matched to the periodic chromatic functions of the achromat and were also made periodic in the straight sections. In addition, the contributions to the natural chromaticity of the straights were minimized to keep the overcompensation of the chromaticity by the sextupoles in the achromat small.




\subsection{Numerical lattice optimization}

\begin{figure}[t]
 \includegraphics[width=\linewidth]{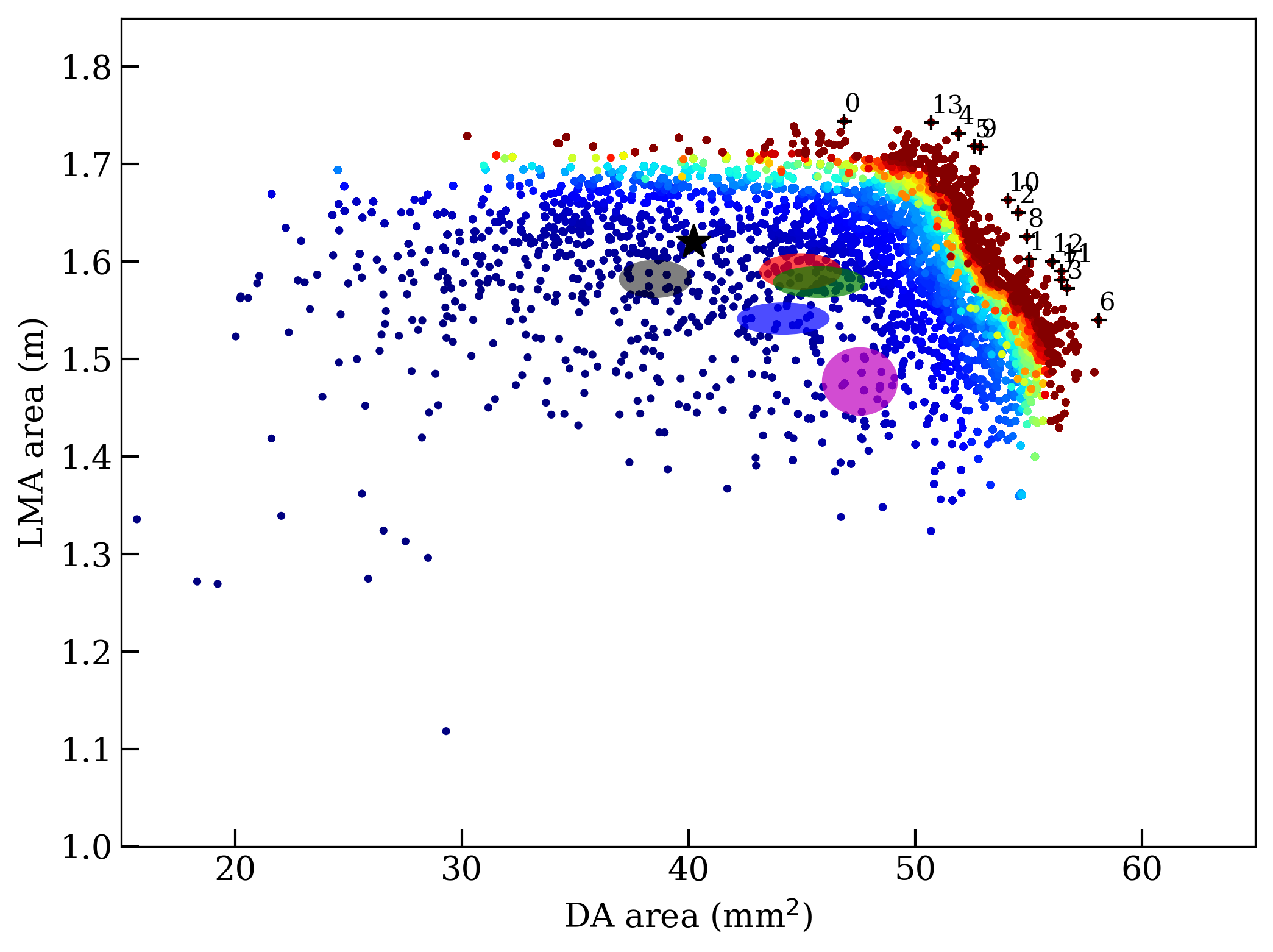}
 \caption{ \label{fig:pareto} Pareto front after 150 generations for a H6BA cell without left/right symmetry of nonlinear elements. Ellipses represent ensemble evaluation  of solutions from the Pareto front with 40 error seeds and 500 turn tracking. }
 \end{figure}

Multi-objective genetic optimization (MOGA) algorithms, in particular NSGA-II \cite{nsga}, have been first applied to storage ring design two decades ago \cite{Emery:2005ju} and since then have found widespread application to storage ring beam dynamics optimization, becoming a standard approach in the light source community.
Due to the number of lattice elements and magnet families, application of MOGA to PETRA~IV has been rather challenging. While 
the analytical optimization procedure described previously leaves no free parameters for numerical optimization of linear optics, lifting those constraints and performing a purely numerical search with approximately a hundred quadrupole families turns out computationally prohibitive. Moreover, calculations without taking the errors into account greatly overestimate the dynamic aperture and the momentum acceptance, and the optimization has to be performed on a statistically significant ensemble of machine realizations with errors.  
Optimization is computationally possible when limited only to the nonlinear elements, i.e. the sextupoles and the octupoles. An example of optimization with six sextupole families and four octupole families per cell is shown in Figure~\ref{fig:pareto}. 

For the optimization with NSGA-II the multi-objective optimization framework \textit{pymoo}~\cite{pymoo} has been used together with \textit{Elegant}~\cite{elegant}. Two objective has been used for MOGA: The area of the upper half of the dynamic aperture for on-momentum particles and the area of the local momentum acceptance of one H6BA cell. The chromaticities were constrained to be in the range of $5 < \xi < 9$. Tracking for DA and MA was done with RF cavities and synchrotron radiation on for 500 turns and without aperture limitations. 

Optimization of both the dynamic aperture and the momentum acceptance requires taking errors into account. Without errors, solutions whose tune footprint crosses both integer and half-integer resonances could be found. With sufficiently small beta-beating (few percent) half-integer resonance crossing appears to be not harmful~\cite{Jiao:2017,Wang:2017}, while the integer resonance always limits the acceptance. With a larger beta-beating both the integer and half-integer resonances limit the acceptance. 


In the optimization procedure random relative gradient errors of quadrupoles and sextupoles with rms. of $2\times 10^{-4}$ and tilt errors with rms. $5 \times 10^{-4}$ rad were used to simulate a well corrected lattice with a small beta beating of 1--2\%.

\begin{figure}
\subfloat[\label{fig:tune_shift_x_moga} Tune shifts with horizontal position $x$.]{\includegraphics[width=\linewidth]{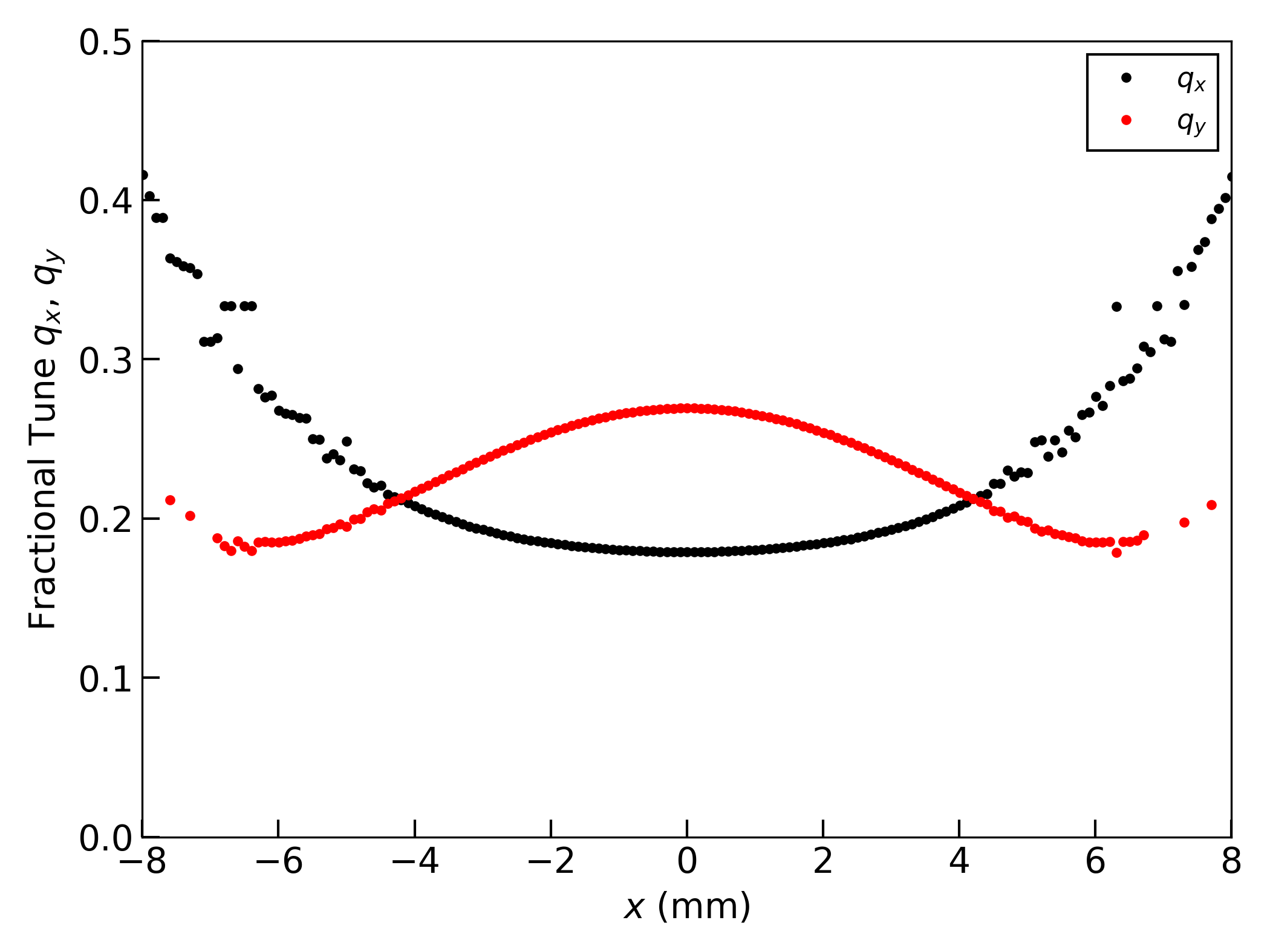}}\\
\subfloat[\label{fig:tune_shift_y_moga} Tune shifts with vertical position $y$.]{\includegraphics[width=\linewidth]{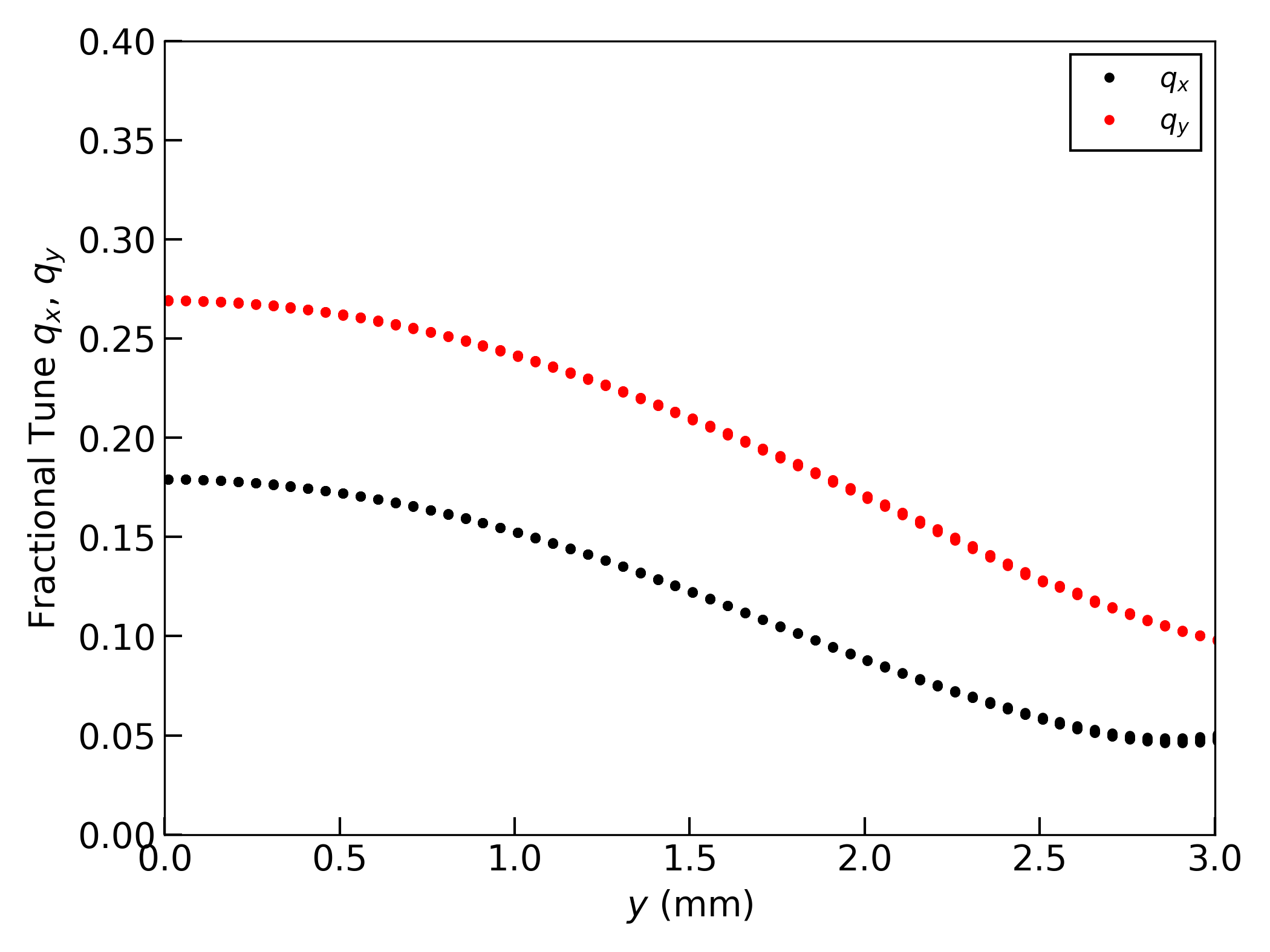}}\\
\subfloat[\label{fig:tune_shift_dp_moga} Tune shifts with relative momentum deviation $\delta$.]{\includegraphics[width=\linewidth]{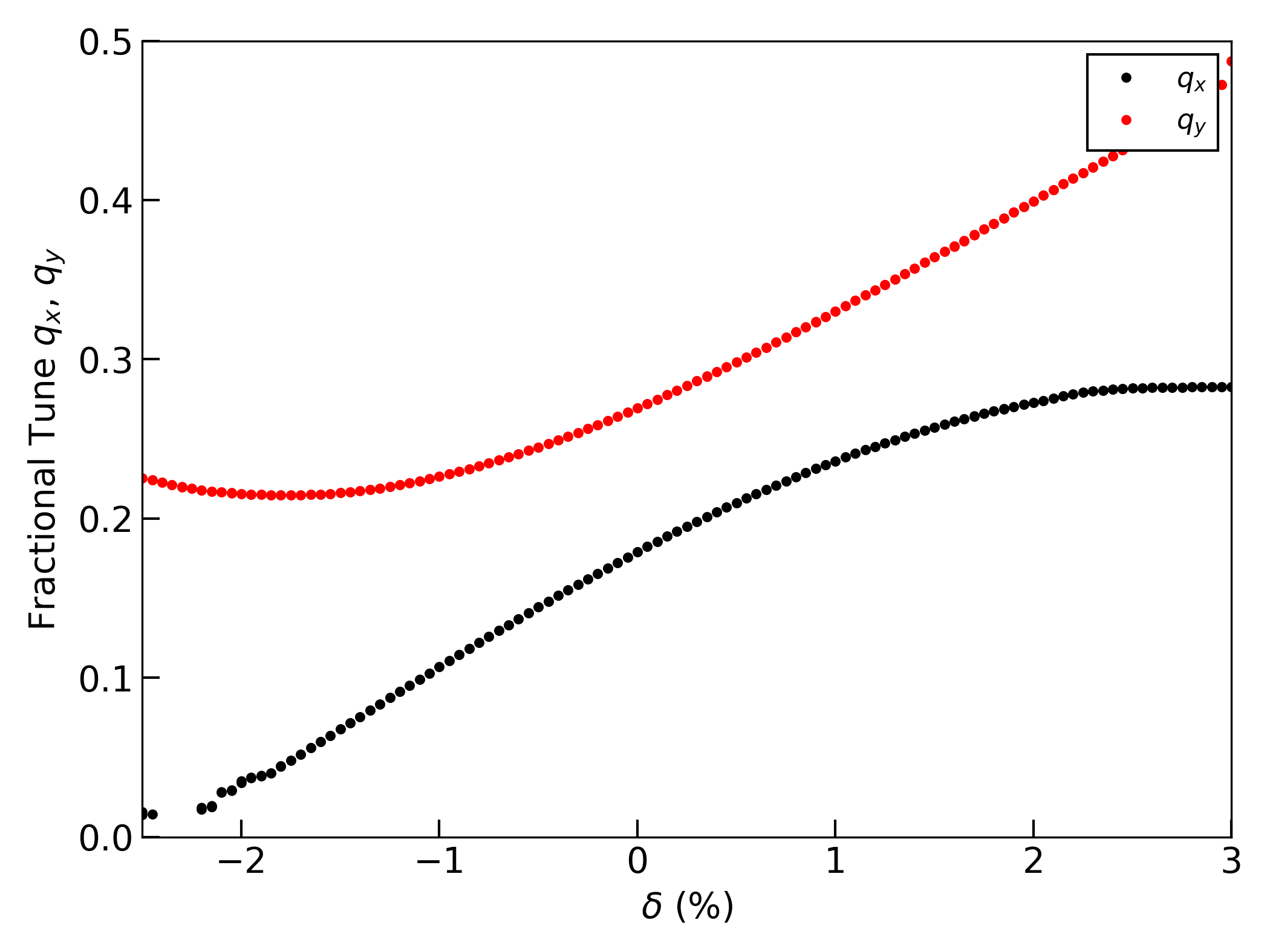}}
\caption{\label{fig:tunefp_moga}Tune shifts for the horizontal (black) and vertical (red) plane after running MOGA. Calculation is done at the injection location $\beta_x=$46 m, $\beta_y=$7 m.}
\end{figure}

The tune shift with horizontal position $x$, vertical position $y$ and relative momentum deviation $\delta$ after running MOGA for a solution with a compromise of DA and LMA is shown in Fig.~\ref{fig:tunefp_moga}.
 
\begin{figure*}
\subfloat[\label{fig:fm_xy_4d} Diffusion map for on-momentum particles and different initial transverse offsets $(x,y)$ without RF cavities and synchrotron radiation (4D).]{\includegraphics[width=0.495\linewidth]{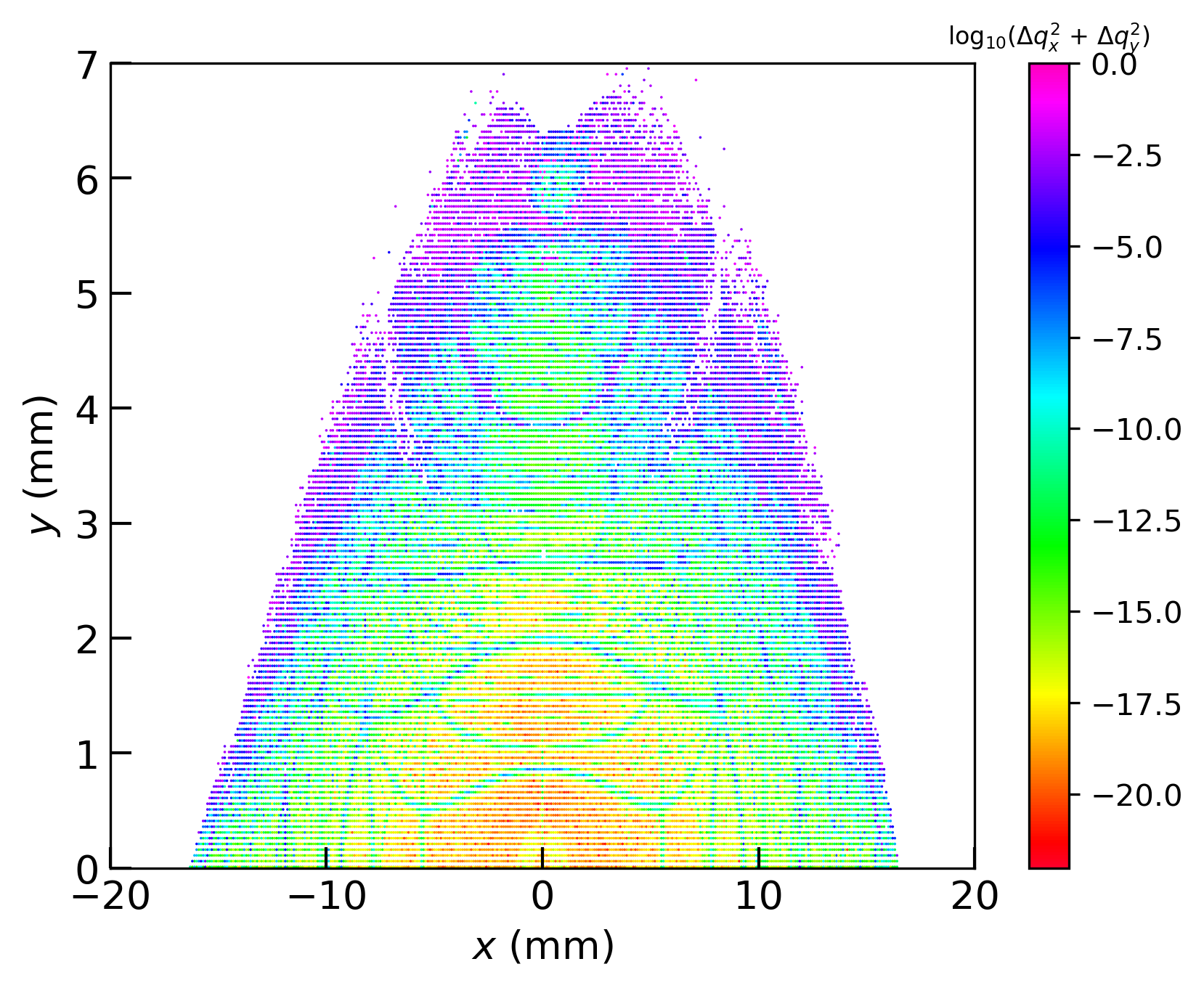}}
\hfill%
\subfloat[\label{fig:fm_qxqy_4d} Fractional tune $q_{x,y}$ footprint for on-momentum particles and different initial transverse offsets $(x,y)$ without RF cavities and synchrotron radiation (4D).]{\includegraphics[width=0.495\linewidth]{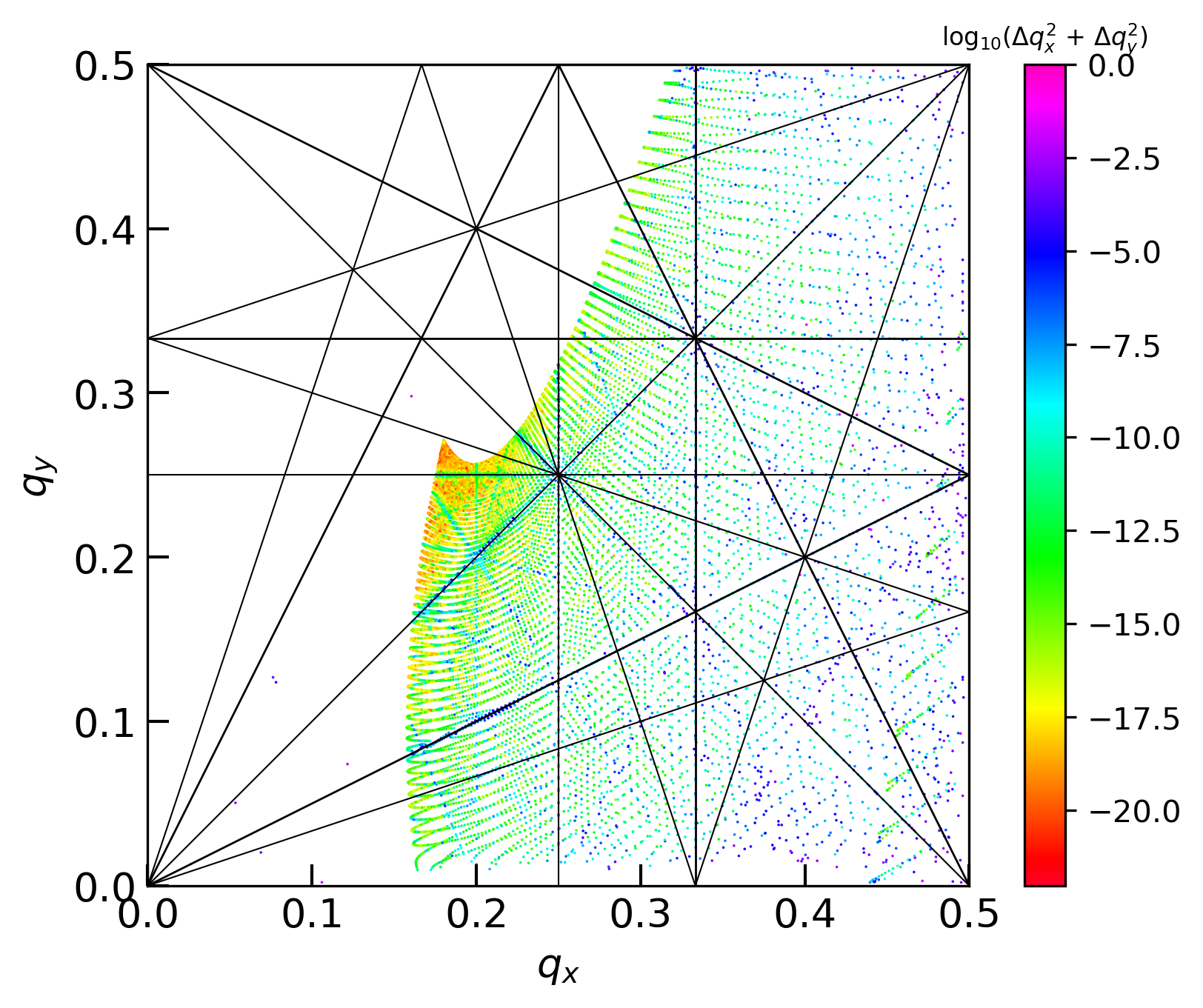}}\\
\subfloat[\label{fig:fm_xy_6d} Diffusion map for on-momentum particles and different initial transverse offsets $(x,y)$ with RF cavities on and synchrotron radiation (6D).]{\includegraphics[width=0.495\linewidth]{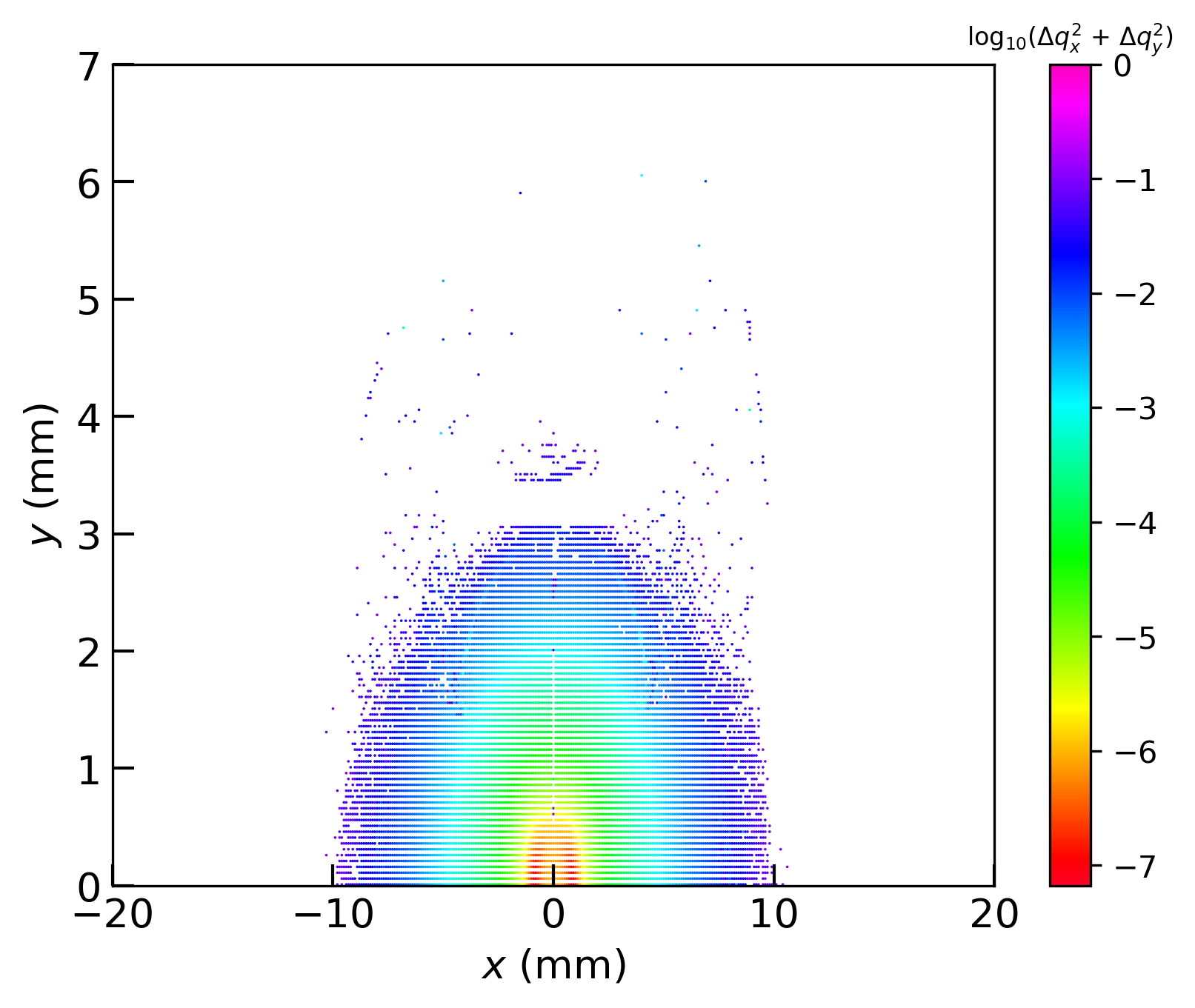}}
\hfill%
\subfloat[\label{fig:fm_qxqy_6d} Fractional tune $q_{x,y}$ footprint for on-momentum particles and different initial transverse offsets $(x,y)$ with RF cavities on and synchrotron radiation (6D).]{\includegraphics[width=0.495\linewidth]{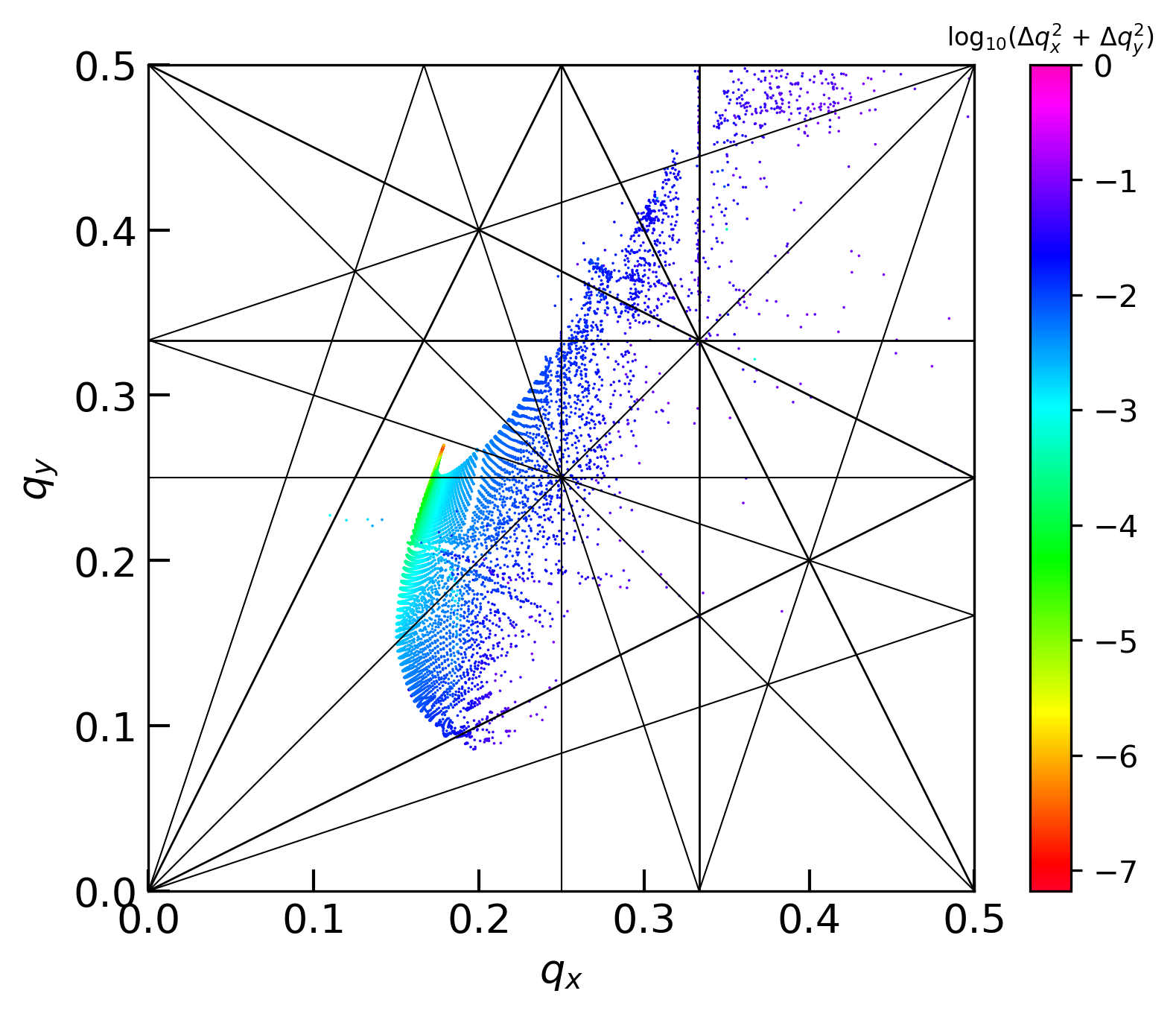}}
\caption{ \label{fig:fma_xy} Frequency map analysis in the transverse space. Calculations are done at the injection location $\beta_x=$46 m, $\beta_y=$7 m.}
\end{figure*}

The frequency map \cite{Laskar:2003zz} in $(x,y)$-space for $\delta=0$ is shown in Fig.~\ref{fig:fm_xy_4d}\ref{fig:fm_xy_6d}. The color code represents the tune diffusion $d=\log_{10}\left( \Delta q_x^2 + \Delta q_y^2\right)$ where $\Delta q_{x,y}$ are the differences in horizontal and vertical tunes from the first and the second half of the tracking, computed over 1000 turns. Note that there is a substantial differences in the dynamic aperture and the frequency maps computed with and without including synchrotron motion into account, which is primarily due to the excitation of synchrotron oscillations through the path lengthening, which is proportional to the transverse particle action \cite{Emery:1992tx}.

\begin{figure*}
\subfloat[\label{fig:fm_xd_4d} Diffusion map for different initial horizontal offsets and off-momentum particles $(x, \delta)$ without RF cavities and synchrotron radiation (4D).]{\includegraphics[width=0.495\linewidth]{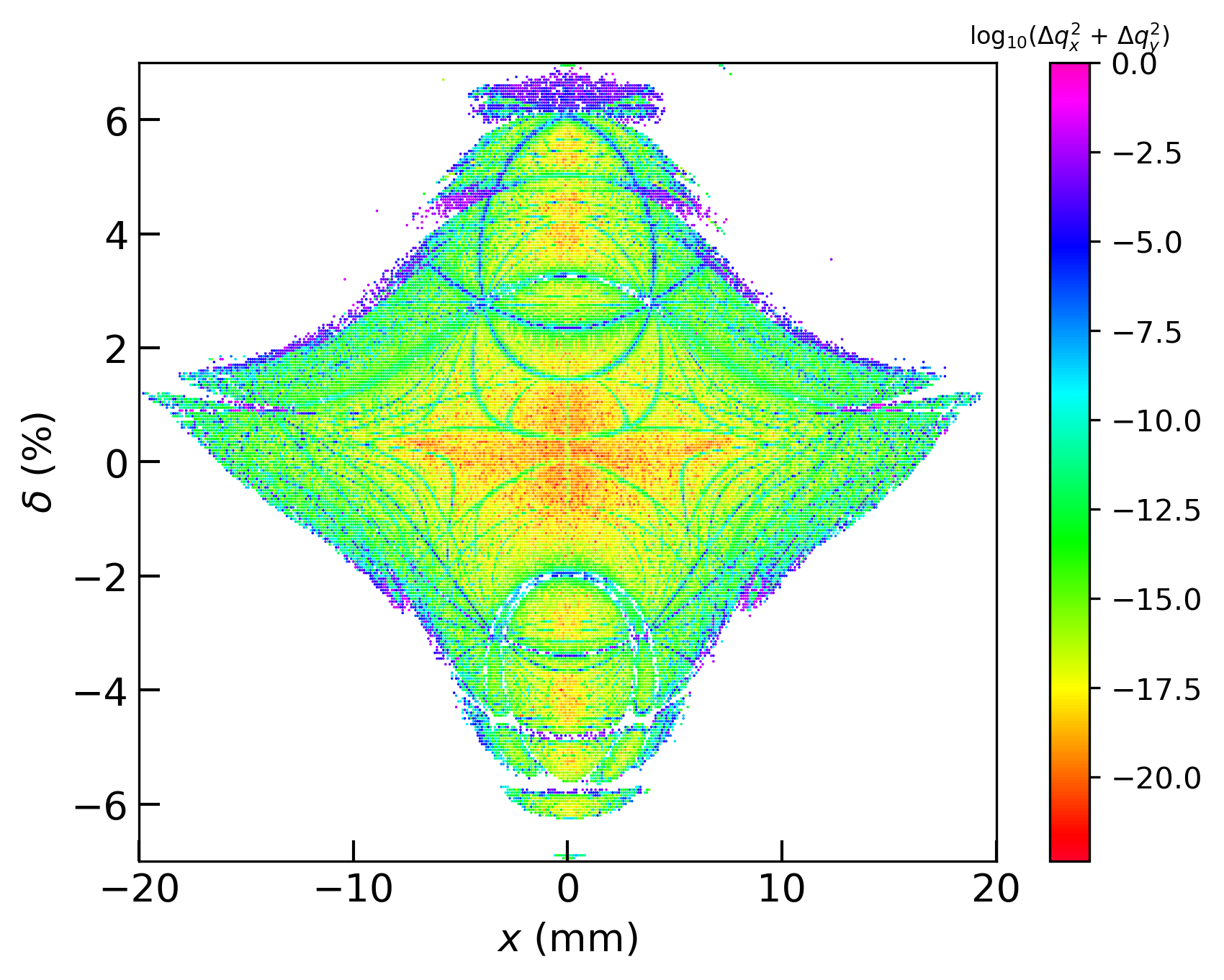}}
\hfill%
\subfloat[\label{fig:fm_qxqy_xd_4d} Fractional tune $q_{x,y}$ footprint for different initial horizontal offsets and off-momentum particles $(x, \delta)$ without RF cavities  and without synchrotron radiation (4D).]{\includegraphics[width=0.495\linewidth]{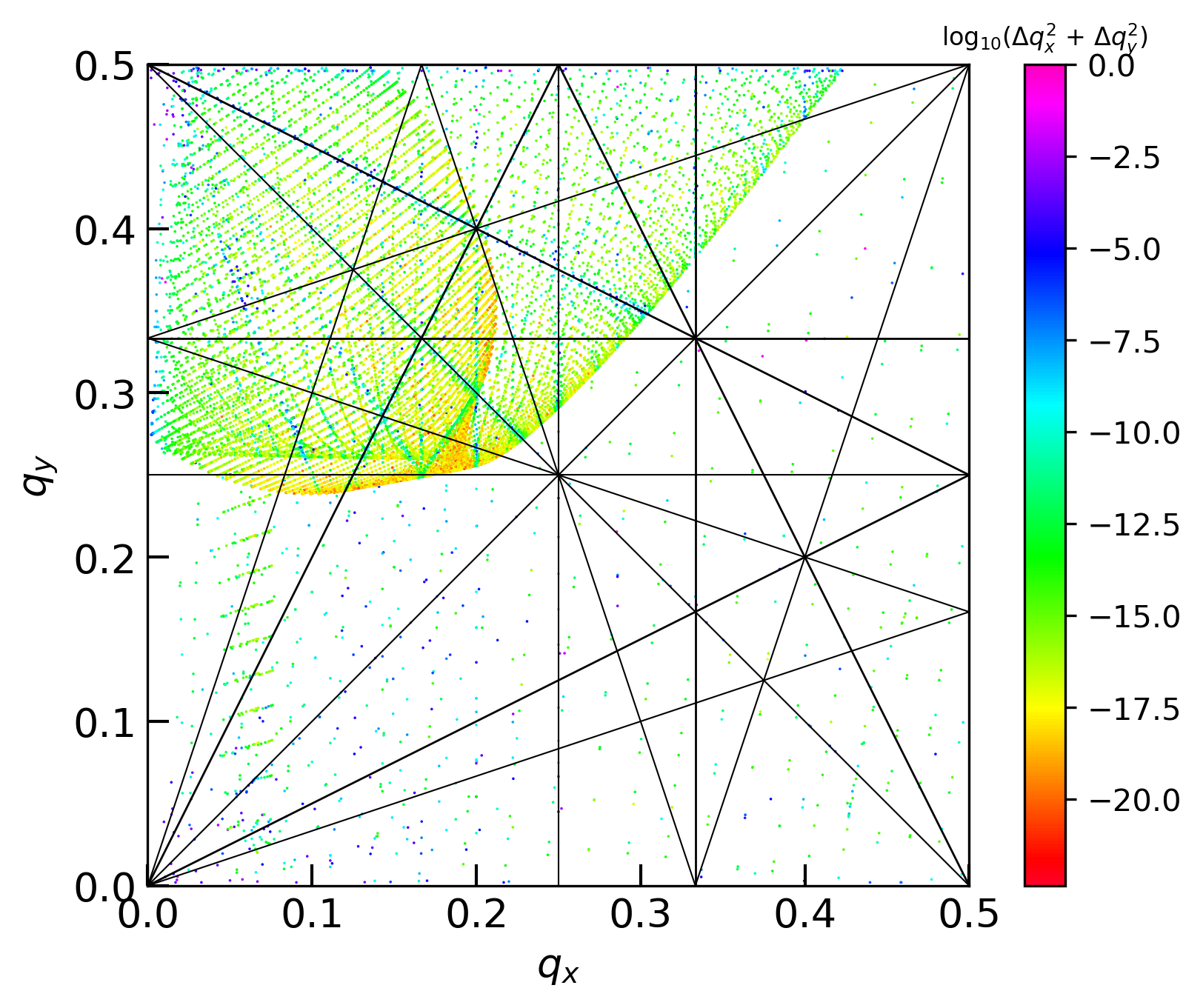}}\\
\subfloat[\label{fig:fm_xd_6d} Diffusion map for different initial horizontal offsets and off-momentum particles $(x, \delta)$ without RF cavities on and synchrotron radiation (6D).]{\includegraphics[width=0.495\linewidth]{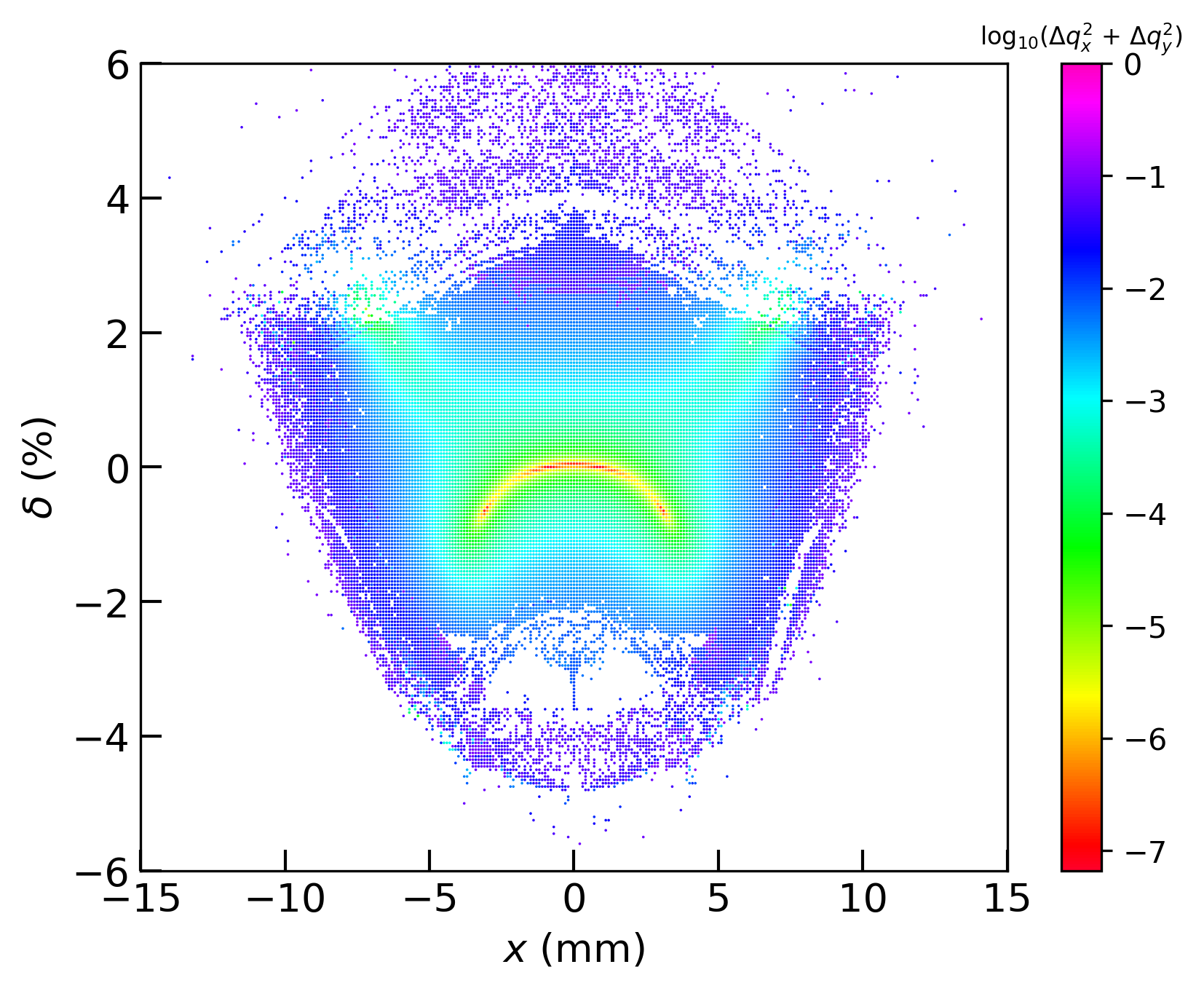}}
\hfill%
\subfloat[\label{fig:fm_qxqy_xd_6d} Fractional tune $q_{x,y}$ footprint for different initial horizontal offsets $(x, \delta)$ and off-momentum particles with RF cavities on and synchrotron radiation (6D).]{\includegraphics[width=0.495\linewidth]{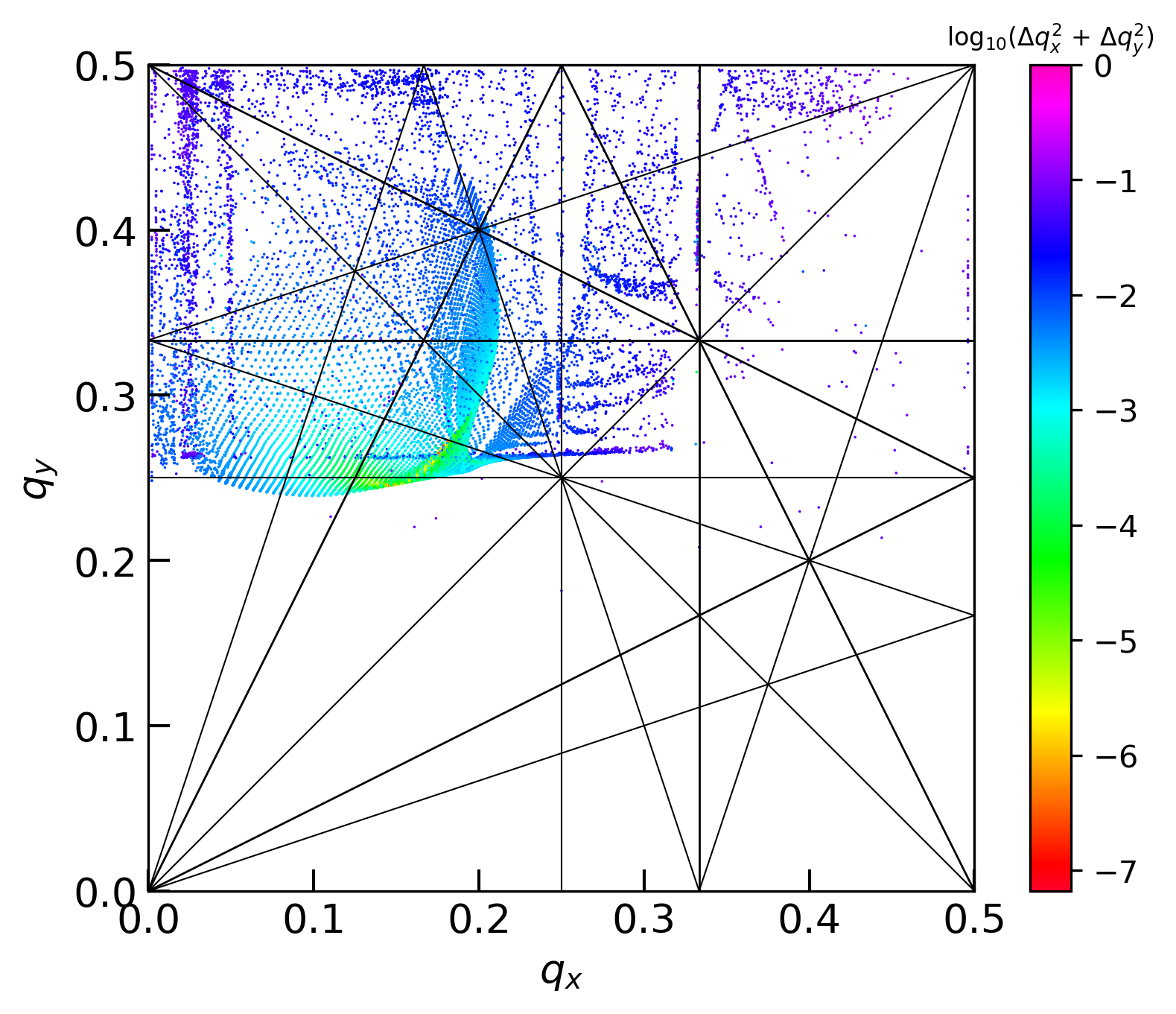}}
\caption{ \label{fig:fma_delta} Frequency map analysis in the momentum space.
Calculations are done at the injection location $\beta_x=$46 m, $\beta_y=$7 m.}
\end{figure*}

The frequency map in $(x,\delta)$-space for $y=0$ is shown in Fig.~\ref{fig:fm_xd_6d}. The color code represents the tune diffusion over 1000 turns.

\subsection{Tolerances}\label{sec:tolerance}

Alignment errors and multipole errors in magnets are usual sources of machine imperfection. 
While the allowed multipole errors are not dissimilar to what has been specified and achieved at many accelerator facilities, i.e. at the $5 \times 10^{-4}$ level, the sensitivity to alignment is significantly increased. This is usual for the multi-bend achromat lattices  and is due to the fact that for a large number of uncorrelated quadrupole offsets the orbit error scales roughly as $\Delta_{x,t} \propto \sqrt{N_{MAG}} B'$ where $N_{MAG}$ is the number of magnets and $B'$ the typical magnetic field gradient. Large offsets in sextupoles in turn have a feed-down effect on linear optics and the orbit becomes unstable for alignment errors in the 5 $\mu m$ range. Such alignment precision cannot be realistically achieved and a certain ``bootstrapping'' procedure should be applied in order to establish circulating beam. This procedure follows standard approach used in MBA lattice commissioning (see e.g. \cite{SajaevComm}). For PETRA IV the commissioning toolbox \cite{Hellert:2019gfx,Hellert:2022zmj} is used  and the procedure consists of the following steps: first the beam is "threaded" through the machine by applying successive trajectory correction until the beam makes several full turns. Trajectory BBA is then applied, and the trajectory again corrected. Sextupoles are ramped up, interleaved with trajectory correction. At this stage the dynamic aperture and momentum acceptance are sufficient to provide beam accumulation and perform optics measurement. We assume that at this stage of the machine commissioning a BBA procedure can be utilized to reduce the BPM offset to 30~$\mu$m. A future version of the commissioning simulation will include these studies explicitly. For now, the BPM offsets are artificially reduced and orbit feedback is applied. After linear optics measurement and correction with the LOCO \cite{SAFRANEK199727} method, the beam and lattice parameters are evaluated.
While a full start-to-end commissioning procedure for PETRA~IV has yet to be developed, error analysis based on a simplified chain as described above is a good approximation to establish error tolerances. 

\begin{figure}
\includegraphics[width=.9\linewidth]{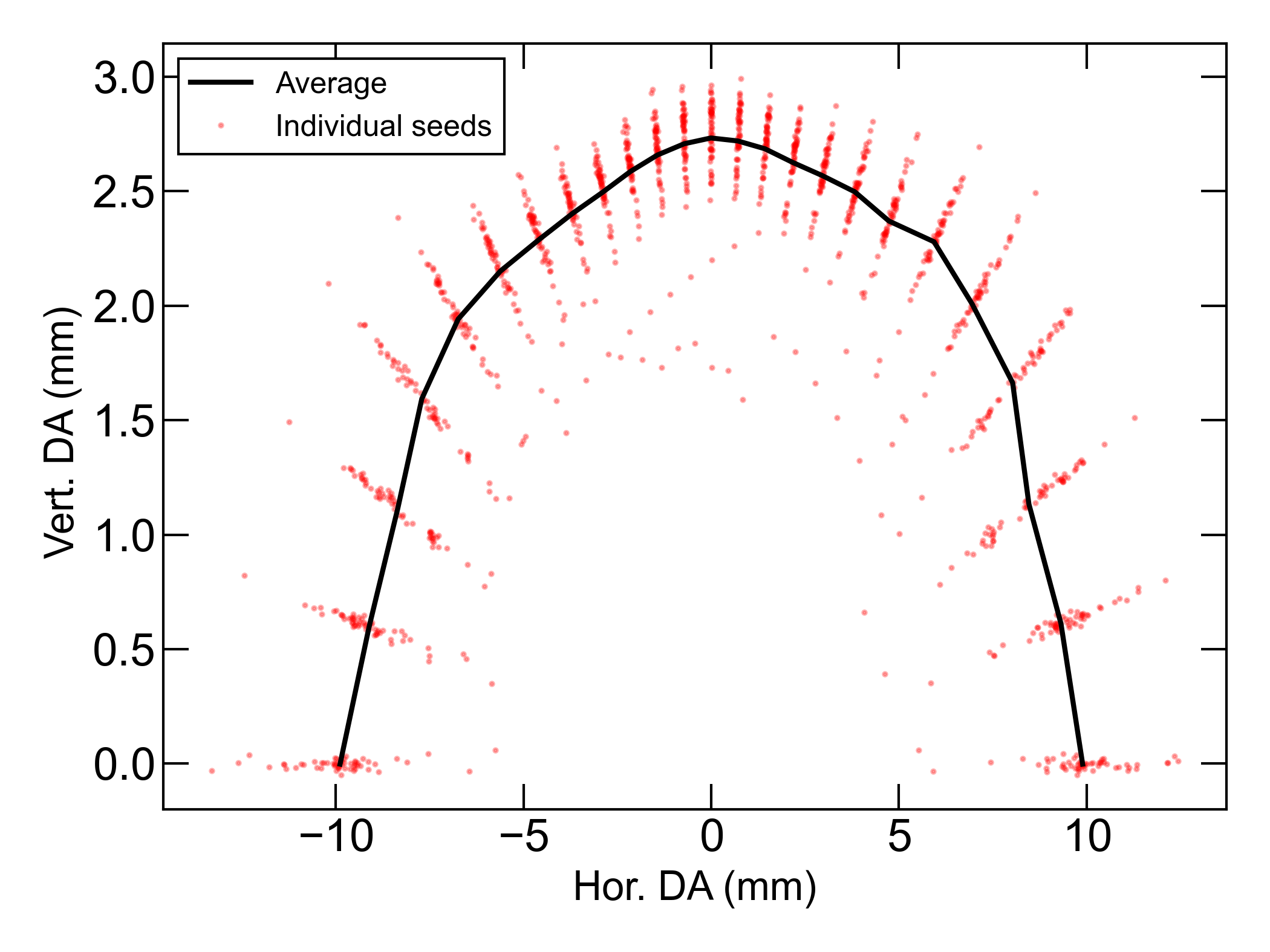}
\includegraphics[width=.9\linewidth]{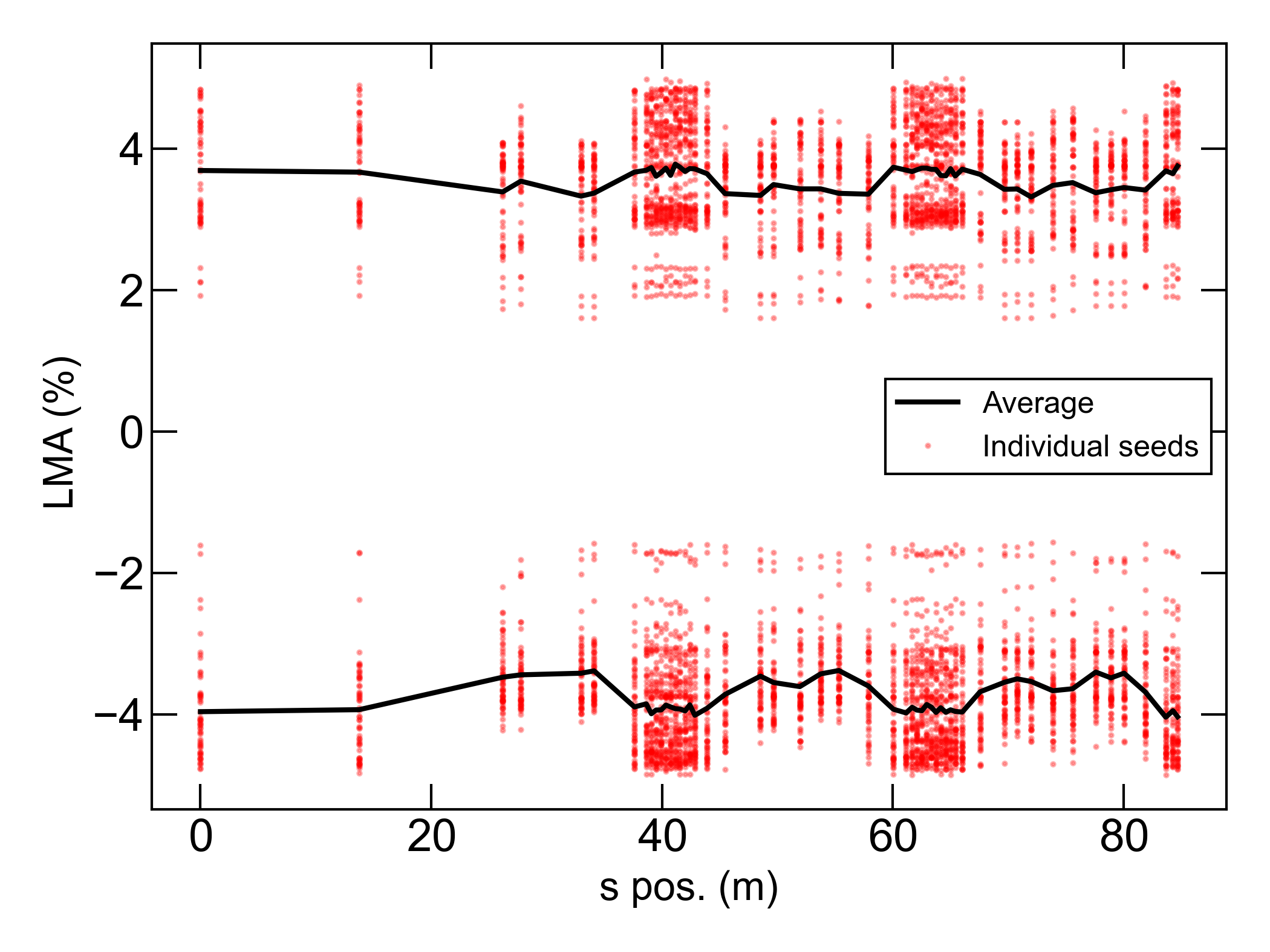}
\caption{\label{fig_da_loco} Dynamic aperture (top) and local momentum acceptance (bottom) after optics correction for 50 error seeds. The black line indicates the average value.}
\end{figure}

The alignment and field error tolerances were chosen based on statistical evaluation of various error scenarios, additionally allowing for some safety margin. At the same time, it was assured that these values are also achievable in practice. The summary of error tolerances are given in Table \ref{tab:tolerances}. Results of the dynamic aperture and local momentum acceptance after the optics correction are shown in Figure \ref{fig_da_loco}. Recently a python version of the commissioning simulation toolbox has been developed and tested \cite{Assmann:2023zku}.

\begin{table}[h]
\caption{\label{tab:tolerances} Assumed alignment errors }
\begin{ruledtabular}
\begin{tabular}{lrrr}
Alignment error & Symbol & Value (rms.) & Unit \\
\hline
Magnet hor., vert., long. & $\Delta x$,$\Delta y$,$\Delta s$  & 30, 30, 100& $\mu$m\\
Magnet roll, pitch yaw & $\Delta \phi$  & 100 & $\mu$rad  \\
Girder hor., vert., long. & $\Delta x$,$\Delta y$,$\Delta s$  & 100, 100, 100 & $\mu$m\\
Girder roll, pitch yaw & $\Delta \phi$ & 100 & $\mu$rad  \\
BPM hor., vert. & $\Delta x$,$\Delta y$  & 500, 500 & $\mu$m  \\

\end{tabular}
\end{ruledtabular}
\end{table}

\begin{figure}[t]
\centering
\includegraphics[width=0.9\linewidth]{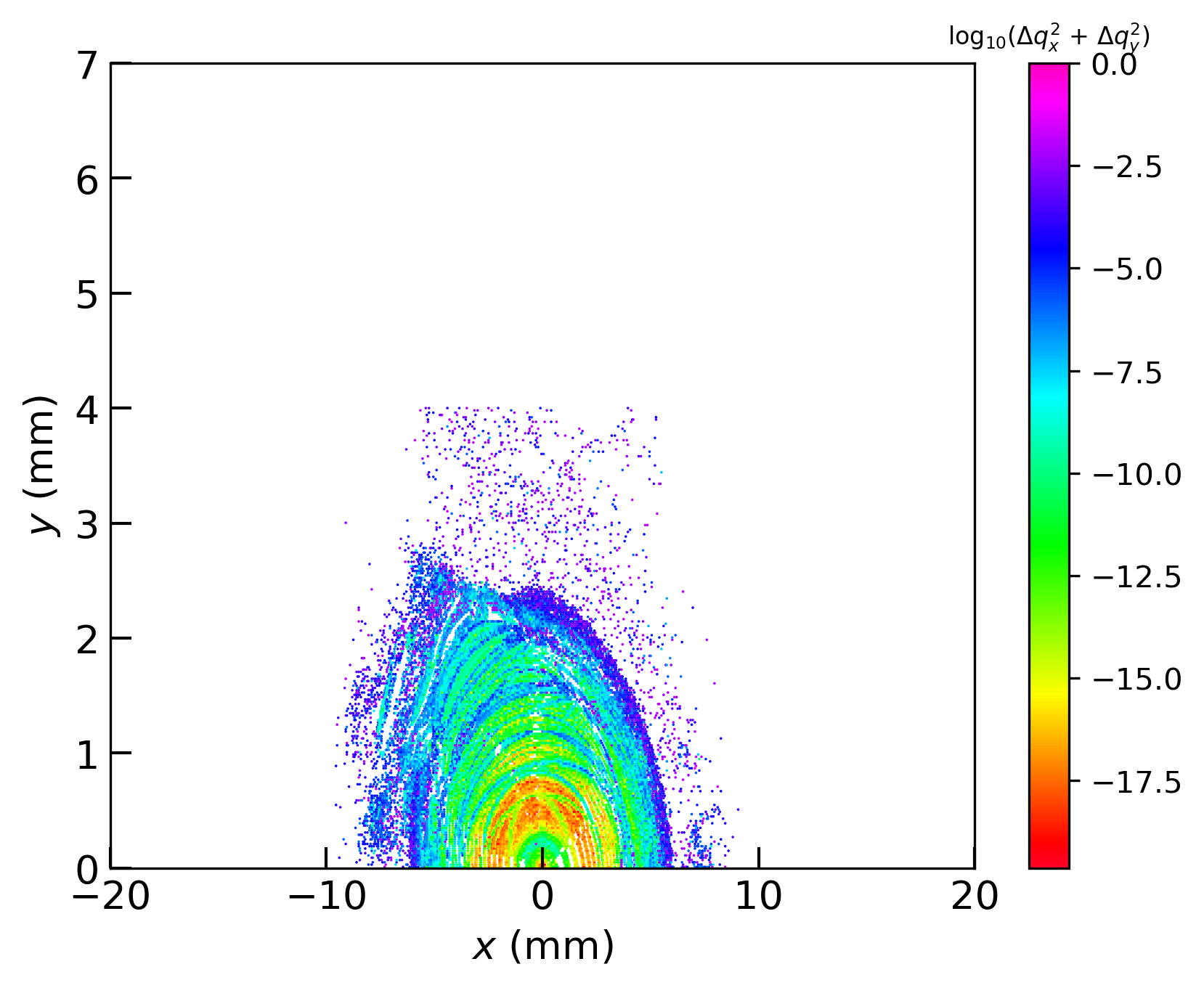}\\
\includegraphics[width=0.9\linewidth]{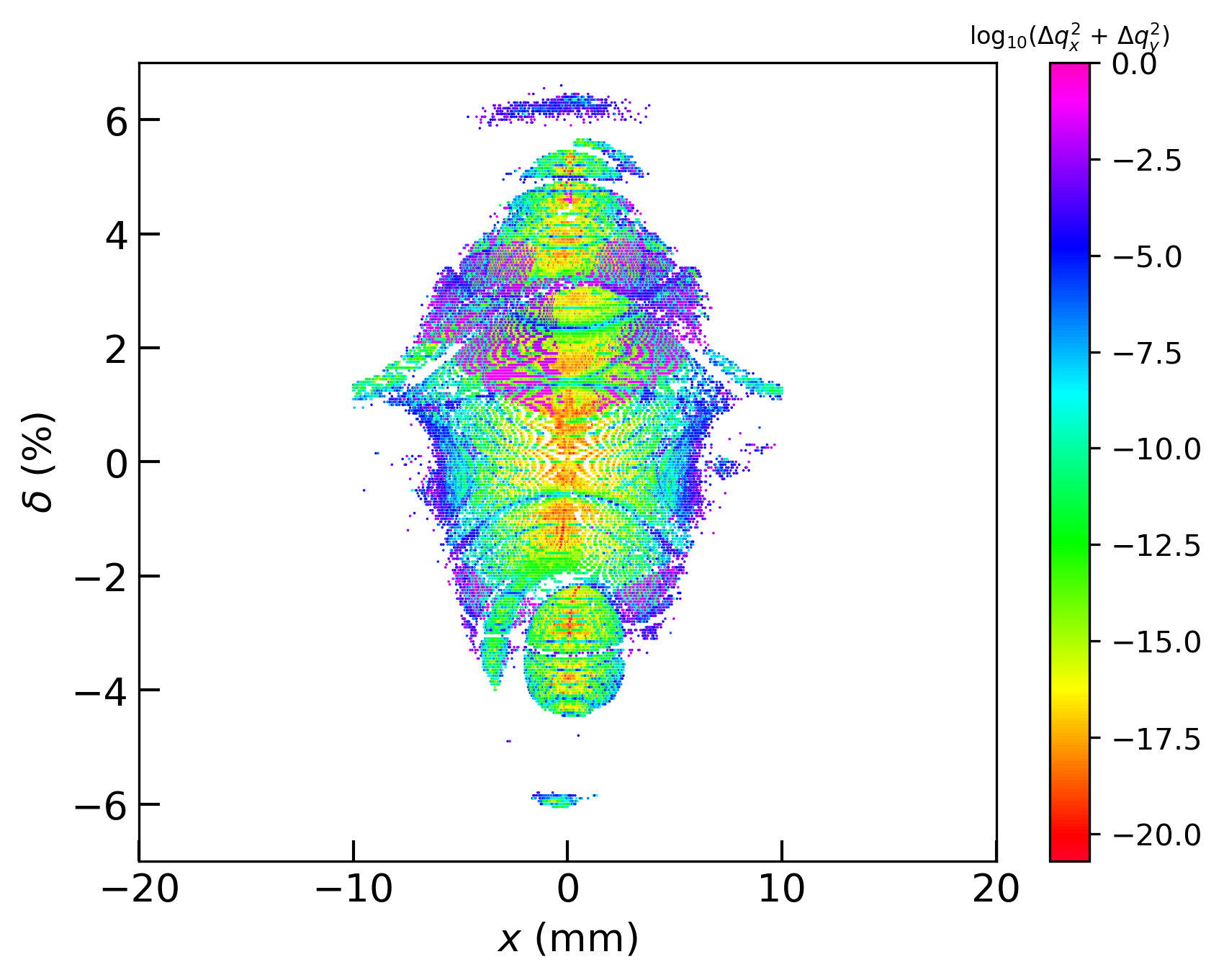}
\caption{ \label{fig:fma_err} FMA with errors. Diffusion map in $(x,y)$ (top) and $(x, \delta)$ (bottom) spaces without RF cavities and synchrotron radiation (4D).}
\end{figure}

While the commissioning simulation procedures serves as the basis for tolerance specification, insight into achievable machine performance can be gained via frequency map analysis with a simplified error model. Reduced errors are introduced so that the beta beating and the orbit errors are similar to those expected during machine operation, and only tune and orbit correction performed. An example of frequency map analysis for one of such error seeds is shown in Figure \ref{fig:fma_err}. By comparing it with Figures \ref{fig:fma_xy},\ref{fig:fma_delta},\ref{fig:tune_shift_dp_moga} one sees that the resonances visible in the ideal frequency map, most notably the integer and the half-integer, limit the momentum acceptance and the dynamic aperture under influence of errors.

\subsection{Collimation}

Since the beam energies and densities in third-generation light sources are typically not sufficient to cause significant material damage, the collimation system design for such machines has not been considered particularly challenging. Large beam losses outside of collimators and dumping the beam on the beam pipe apertures was usually accepted. Experience at PETRA III showed that even if the mechanical material damage has indeed not been observed, beam losses can cause demagnetization of material in the insertion devices \cite{Sahoo:2015xbu}, \cite{Vagin:2014csa}. Analysis of beam losses for forth-generation light sources \cite{Dooling:2022uld} suggests that much more attention has to be payed to the collimation system design due to several orders of magnitude higher energy density in the low-emittance beam. Moreover, increased use of permanent magnets in the accelerator lattices puts more stringent requirements on the radiation environment. 

The collimation system for PETRA IV \cite{Jebramcik:2022nyc} was designed  with the following objectives:
intercepting injection losses and protecting the machine against beam steering errors at injection;
intercepting Touschek and gas-scattered particles;
handling emergency and routine beam dump scenarios.

The system consists of horizontal collimators placed in the downstream dispersion bump of the four of the H6BA cells (see Fig.~\ref{fig_cell}), and of two vertical collimators placed in one of the long straight sections (South). The horizontal collimators are at the locations of maximum dispersion, and thus also serve as energy collimators.

The interception of losses caused by transverse offset and angle errors at injection can be done with 100 \% efficiency whenever the collimator gaps are smaller than the dynamic aperture at the collimator locations (see \cite{Jebramcik:2022nyc}). The energy errors at injection also cause the beam to be lost only at the position of the horizontal collimators.

\begin{figure}
\includegraphics[width=.99\linewidth]{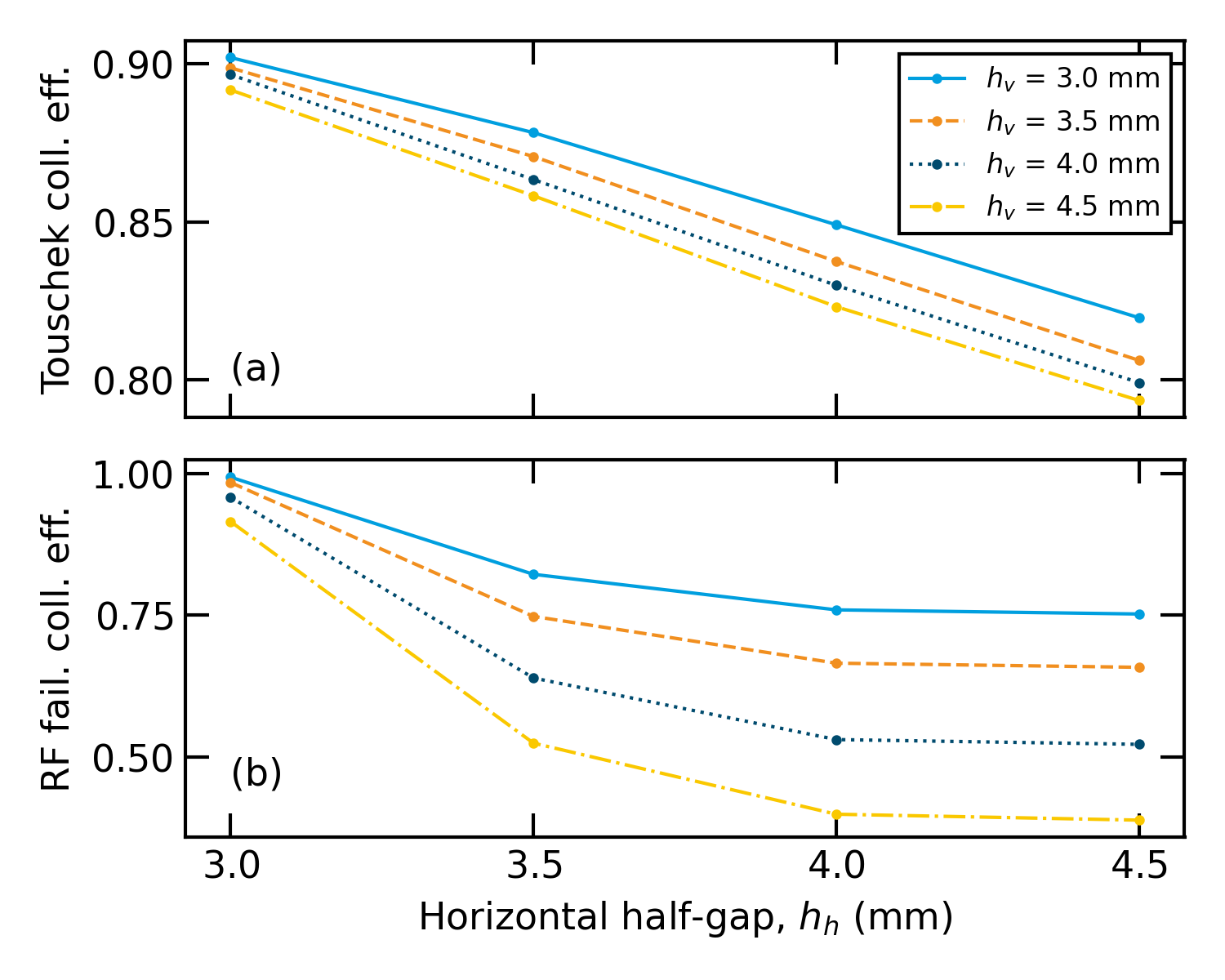}
\caption{\label{fig_collimation_1} Touschek collimation efficiency (a) and efficiency of intercepting the beam after switching off the RF voltage (b) as a function of horizontal $h_h$ and vertical $h_v$ collimator half-gaps.}
\end{figure}

Collimation of Touschek-scattered particles is more complicated. Such particles are lost when their energy after scattering exceeds the momentum acceptance. For large energy deviations where the motion is unstable the linear lattice functions can not be used to predict the particle trajectories and the losses are generally not limited to the positions with the smallest acceptance of the on-energy optics. 
Moreover, the scaling of any multi-bend achromat cell such as the H6BA to a machine of large circumference results in a small value of the dispersion function, while the maximum value of the beta function only depends on the cell length and remains similar to the one for smaller rings with a similar lattice type and cell length. This results in either a need for smaller collimator gap  or poorer efficiency of energy collimation compared to other similar multi-bend achromat lattices of smaller size such as ESRF-EBS or APS-U. 
Extensive tracking studies have been performed with {\it Elegant} \cite{elegant} to understand the beam loss patterns of Touschek-scattered particles. Both the ideal optics and a statistical ensemble of perturbed optics with 5 \% beta beating were used. Figure \ref{fig_collimation_1}(a) shows the average total Touschek collimation efficiency as a function of vertical and horizontal collimator gap settings. Below approximately 3 mm the collimator gaps would strongly affect the injection efficiency by effectively reducing the dynamic aperture, and start contributing significantly to the overall impedance budget. With the minimum gap of 3 mm, roughly 90 \% Touschek collimation efficiency can be achieved. The remaining losses are approximately equally distributed between insertion devices and the rest of the machine.

 Due to the relative energy loss per turn of approximately $10^{-3}$ the beam reaches the energy acceptance in 100-400 turns after switching off the RF voltage, depending on the assumptions on the optics errors and on the momentum acceptance (see Figure \ref{fig_da_loco}). The loss occurs within 0.7 - 3 ms and is faster than the time during which the field of a main magnet would decay significantly (more than 1 \%) after a power supply failure. The PETRA IV power supply system will have redundancy (the so-called "hot-swap" \cite{Schroer2019}), but an unrecoverable power supply failure can be handled by switching off the RF system and dumping the beam on the horizontal collimators. Figure \ref{fig_collimation_1}(b) shows the efficiency of intercepting the beam by horizontal collimators after the RF power has been switched off as a function of gaps for the set of simulations  on a statistical ensemble of lattices with 5 \% beta beating. Unlike the Touschek-scattered particles where the particle energy jumps suddenly, here the beam loses energy continuously until the momentum acceptance is reached or the dispersive orbit reaches the aperture limitations. Tracking simulations show that a 3 mm gap is sufficient to provide close to 100 \% interception efficiency. 
 
In all beam dump situations the beam energy density is high enough to cause material damage, and a special vertical kicker magnet would be used to blow up the vertical beam size prior to the beam dump.
Full analysis of various failure modes such as malfunctioning of the feedback systems is outside of the scope of this paper. The collimation settings will have to be adjusted in operation based on the effective momentum acceptance.


\subsection{Collective effects}
The collective effects driven by broad-band (geometric and resisitive wall) and narrow-band (cavity) impedances, the Intra-Beam Scattering (IBS), and the Touschek Scattering have been assessed. 

\subsubsection{Modes of operation}
PETRA~IV will need to be operated with several filling patterns. The two baseline filling schemes are a 200~mA Brightness mode with 1920 bunches and a 80~mA Timing mode with 80 bunches. Both schemes utilize uniform bunch spacing. Other, non-baseline patterns are also foreseen, such as a 2~ns Brightness mode, a 1600 bunch Brightness mode (20 bunch trains followed by 4 empty buckets, CDR), a Hybrid pattern \footnote{Here we limit ourselves to filling patterns with bunches of equal charge. A hybrid pattern may also use high charge `guarding' bunches at the ends of the train. } with only $7/8$ of the ring filled, or a 40-bunch Timing mode. Table~\ref{tab:FS} summarises the main features of different bunch patterns. 

\begin{table}[h]
   \centering
   \caption{PETRA~IV baseline modes and some alternative scenarios.}
   \begin{ruledtabular}
   \begin{tabular}{lrrrr}
       Filling & No. bun. & Spacing & Sep. & Current \\
       \hline
           Brightness & 1920 & Uniform & 4~ns & 200~mA\\
           Timing & 80 & Uniform & 96~ns & 80~mA\\
       \hline
           CDR & 1600 & 20b+4e & 4~ns & 200~mA\\
           Hybrid & 1680 & 7/8 & 4~ns & 200~mA\\
           2~ns & 3840 & Uniform & 2~ns & 200~mA\\
           Timing & 40 & Uniform & 192~ns & 80~mA\\
    \end{tabular}
    \end{ruledtabular}
   \label{tab:FS}
\end{table}

\subsubsection{Impedance}
\begin{figure}
    \includegraphics[width=.95\linewidth]{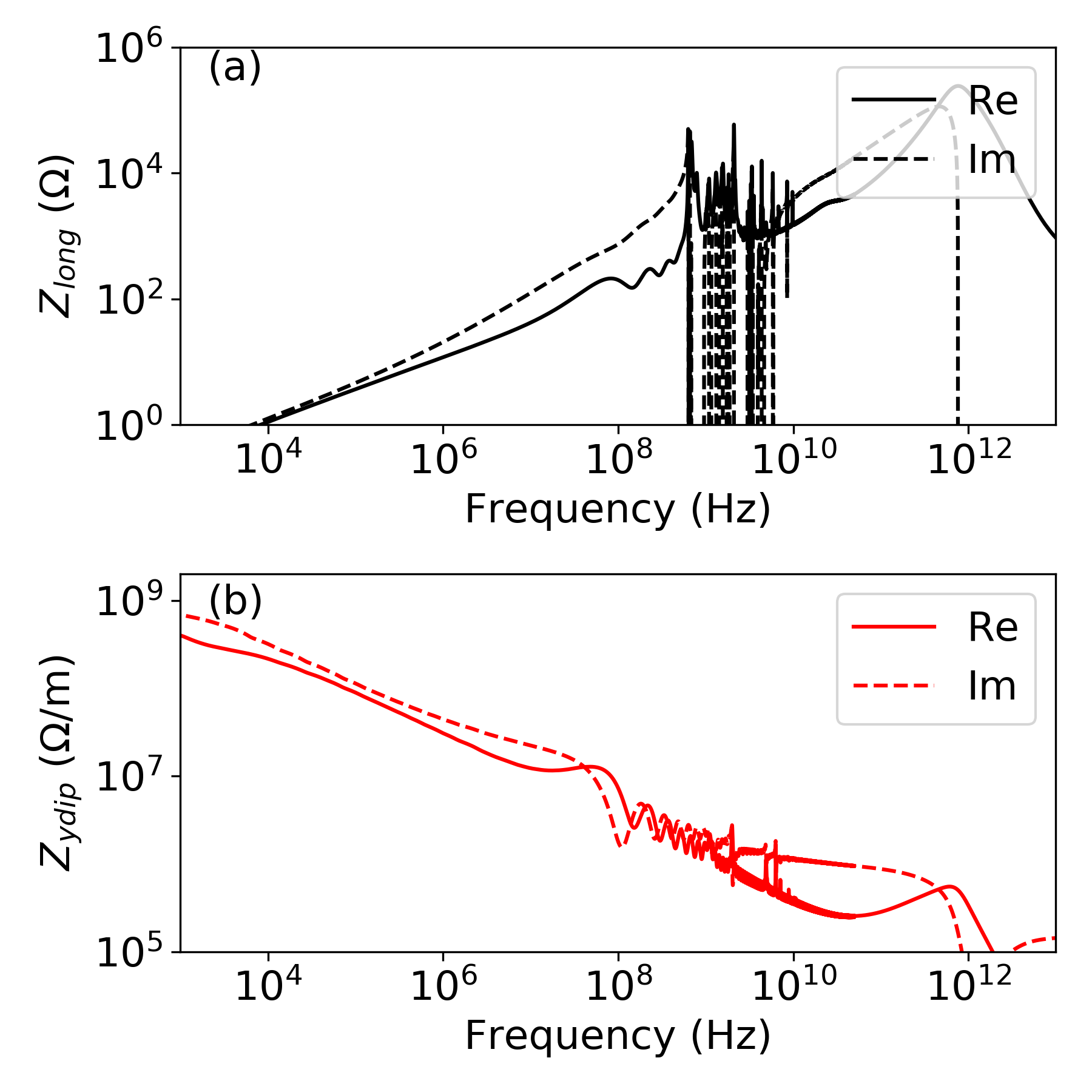}
    \caption{\label{fig:Impedance-comparison} Longitudinal (a) and transverse (b) impedance of the PETRA~IV ring. Solid lines denote the real and dashed -- the imaginary component.}
\end{figure}
The impedance model of the main ring consists of the resistive wall contributions of its standard and ID vacuum chambers and multiple sources of geometric impedance: RF cavities, injection and feedback kickers, BPMs, current monitors, tapers, bellows, and radiation absorbers. The resistive wall contribution is modeled using the IW2D code~{\cite{IW2D}}. The standard vacuum chamber is circular with a 10~mm radius and is made of Cu. The ID chambers are elliptical and made of Al with  a 6~mm vertical full gap; all vacuum chambers are assumed to be coated with NEG (resistivity of 2~$\mu\Omega$m, according to RF measurements of coated tube samples). The coating is not expected to affect the impedance significantly when its thickness does not exceed $\sim~1$~$\mu$m. As for the geometric impedance, at the time of writing of this paper, many hardware components (such as striplines and tapers) are still in the early design stages and they are included in the impedance budget as effective broad-band impedance. According to the CDR studies~\cite{Chae:2019impe}, in practice, we can expect the total geometric impedance to be about or smaller than 0.4~M$\Omega$/m, or less than 30\% of the overall impedance budget~(Table~\ref{tab:imp}).

\begin{table}[h]
\caption{\label{tab:imp} Impedance contributions at chromaticity 6 in the vertical plane.}
\begin{ruledtabular}
\begin{tabular}{lll}
Impedance contribution & Value (M$\Omega$/m) & Share (\%) \\
\hline
RW round chambers & 0.32 & 24 \\
RW ID chambers & 0.60 & 45\\
Geometric impedance&  $\leq 0.4$& $\leq 30$\\
\end{tabular}
\end{ruledtabular}
\end{table}

Due to the small $\beta$-functions at the standard IDs, the transverse resistive wall impedance of the H6BA lattice is also relatively small (Fig.~\ref{fig:Impedance-comparison}). Here we consider a Gaussian bunch with an rms. bunch length of 40 ps, corresponding to the baseline Brightness mode of operation. For such a bunch the effective restive wall impedance is $Z_{y,eff} \approx 0.92$~M$\Omega$/m at chromaticity 6 for the most critical vertical plane. This value is somewhat smaller than the 1.0~M$\Omega$/m figure assumed in the CDR and well within its budget of 1.4~M$\Omega$/m~\cite{Schroer2019}. As will be shown later, the impedance of PETRA~IV allows operating all baseline modes with significant safety margins.

\subsubsection{Single-bunch effects}
\begin{figure}[!htp]
\centering
\includegraphics[width=3.5in]{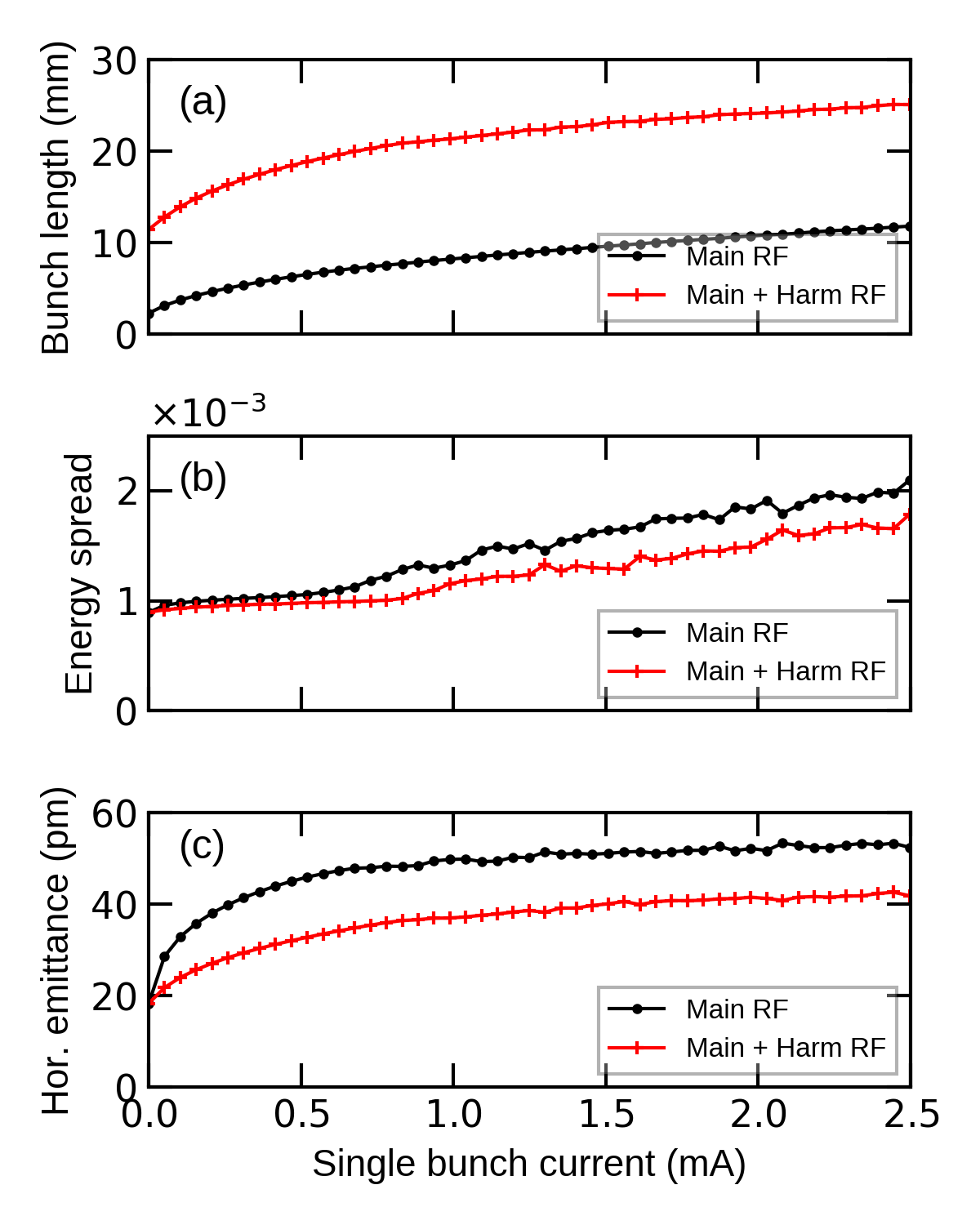} 
\caption{\label{fig:SingleBunchLength}Single bunch length (a), rms. energy spread (b) and horizontal rms. emittance (c) as a function of single bunch current.} 
\end{figure}

\begin{figure}
\centering
    \includegraphics[width=1\linewidth]{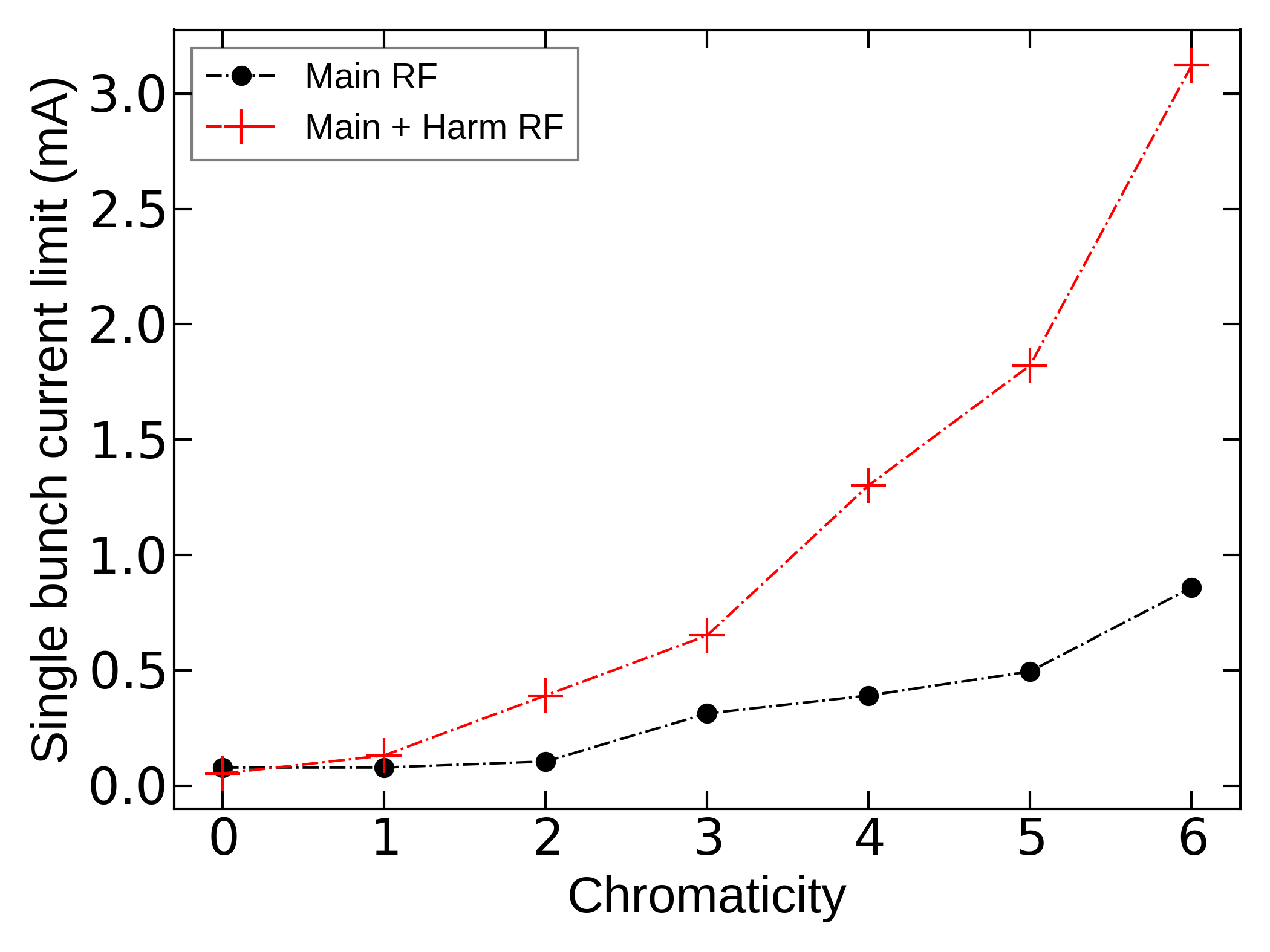}
\caption{ \label{fig:SingleBunchThreshold_b}The single bunch current limit as a function of the chromaticity. } 
\end{figure}

In the longitudinal plane the 3rd order harmonic cavity together with the beam coupling impedance privide significant bunch lengthening. The transverse emittance growth is mostly due to the IBS effect, and is suppressed significantly by the bunch lengthening. Figure~\ref{fig:SingleBunchLength} shows the equilibrium bunch length, rms. energy spread and horizontal rms. emittance as a function of the bunch current for chromaticity 6. Fig.~\ref{fig:SingleBunchThreshold_b} shows the single bunch current limit as a function of chromaticity.
The simulations were performed by particle tracking with {\it Elegant} using $10^5$ macro-particles.
Tab.~\ref{tab:beam_parameters} summarizes the bunch parameters at the target bunch current in different operation schemes.

\subsubsection{Beam-ion effect}
\begin{figure}
    \includegraphics[width=1\linewidth]{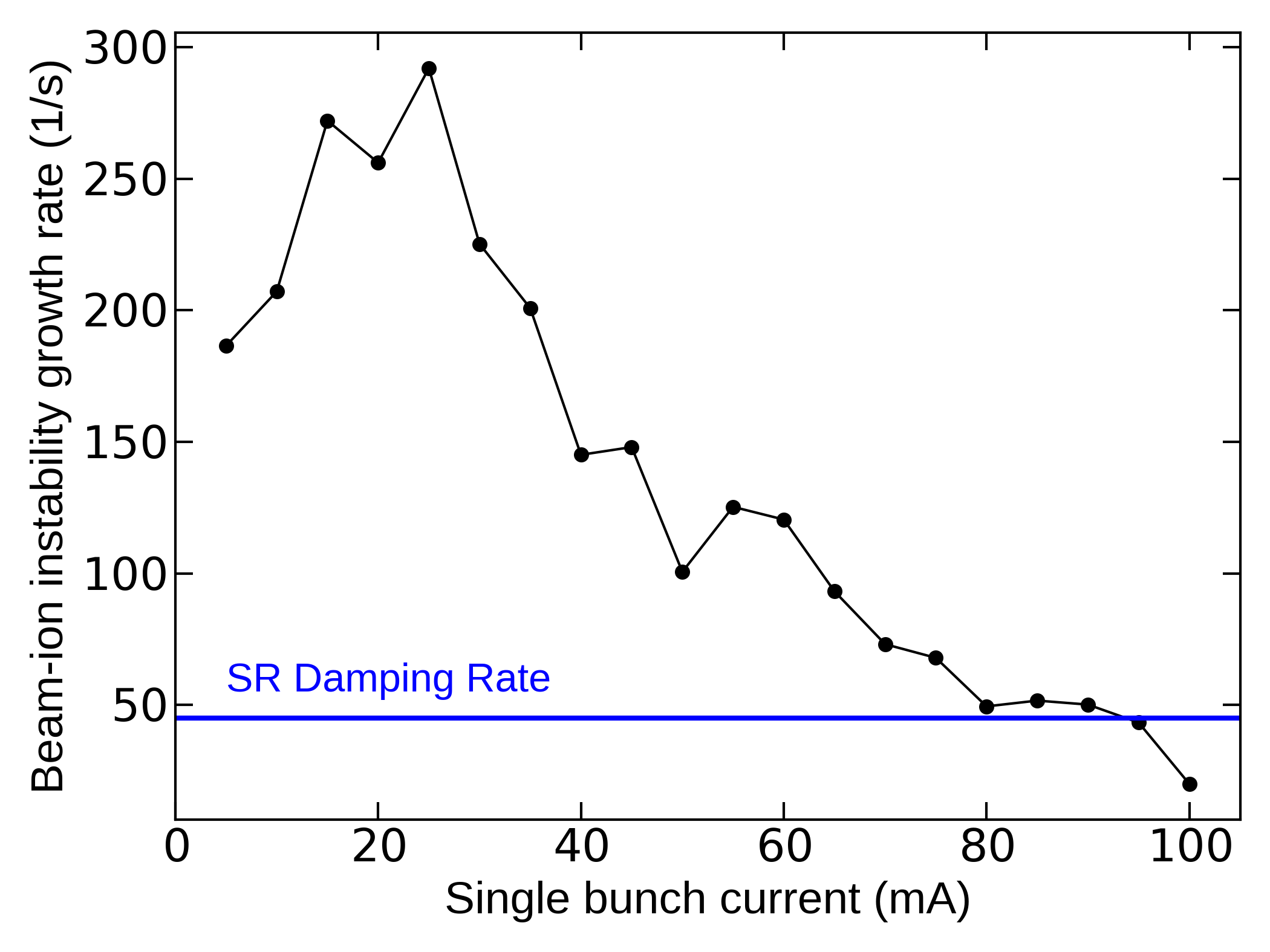}
    \caption{\label{fig:beamion_grwothrate}Beam-ion instability growth rate as a function of total beam current. The ion composition is assumed to be the same APS-U case.}
\end{figure}

 We assessed two types of ion gas composition (see Tab.~\ref{tab:ions}) assuming the total residual gas pressure is 1~nTorr and the gas temperature of 300~K. The filling pattern for the Brightness operation scheme, with 1600 bunches in total. The beam-ion instability ~\cite{PhysRevAccelBeams.23.074401} was simulated using the CETASim~\cite{CETASim} code, which models dipole mode growth with a single averaged beam-ion interaction per turn. The growth rates with a more conservative APS-U gas composition as a function of beam current are shown in Figure~\ref{fig:beamion_grwothrate}. Transverse feedback would be required for emittance stabilization in that case. With the  MAX-IV-type gas composition , the growth rates would be always below the synchrotron radiation damping rate.

\begin{table}[!htp]
\caption{\label{tab:ions} Ion species used in the simulation}
\begin{ruledtabular}
\begin{tabular}{lllll}
Ions &$H_2$ & $CH_4$ & $CO$ & $CO_2$ \\
\hline
APS-U \%  & 43 & 8 & 36 & 13 \\
MAX-IV \% & 96 & 1 & 1 & 2 \\
\end{tabular}
\end{ruledtabular}
\end{table}

\subsubsection{Transverse coupled  bunch instability}
The transverse collective beam stability in the presence of feedback and chromaticity was assessed using the NHT Vlasov solver~\cite{NHT}. This solver treats a complete coupled-bunch problem in a `flat-wake' approximation~\cite{Burov:2013nb}, which is applicable for the wakes dominated by resistive wall effects. It also takes into account the filling pattern: 80 20-bunch trains with a 4~ns bunch separation in case of the Brightness mode, or 80 equidistant bunches in case of the Timing mode. The beam is deemed stable if the growth rate of all coupled bunch modes does not exceed the synchrotron radiation damping rate. 

Without the transverse feedback the instability growth rates exceed the synchrotron damping time of about 3000 turns at chromaticity of 6 for both baseline operation modes: 200~mA Brightness and 80~mA Timing mode (see Table~\ref{tab:gr_rates}). That means that operationally, the beam could be stabilized by the chromaticity setting alone. Yet, for simplifying the commissioning it would be useful if the beam remained stable at low chromaticities close to 0. The growth time at $\xi = 0$ is expected to be larger than 150 turns in all potential filling schemes (Table~\ref{tab:gr_rates}), therefore a transverse feedback with a damping time smaller than 100~turns would be sufficient to stabilize even the most challenging future operational scenarios. Additionally, it has been found that for a machine whose impedance is mainly governed by resistive wall contributions and with a sufficiently strong transverse feedback and chromaticity, irregularities in filling patterns have no significant impact on the transverse beam dynamics \cite{Antipov:2023cfc}.
Thus, with both chromaticity and feedback in place, all operation modes, including non-baseline, are expected to be stable with at least a 100\% safety margin in terms of growth rate.

\begin{table}[h]
\caption{\label{tab:gr_rates} Transverse coupled-bunch instability growth times in the absence of feedback and synchrotron radiation damping for different operation modes. The first two scenarios are baseline, the bottom three are hypothetical.}
\begin{ruledtabular}
\begin{tabular}{llll}
Filling scheme & Current & $\xi = 0$ & $\xi = 6$ \\
\hline
Brightness, 4 ns & 200 mA& 250 turns & $3.9\times10^3$~turns \\
Timing, 80 b. & 80 mA& 770 turns & $2.2\times10^4$~turns \\
\hline
Brightness, 2 ns & 200 mA& 250 turns & $3.4\times10^3$~turns \\
Timing, 40 b. & 80 mA& 640 turns & $1.3\times10^4$ turns \\
Timing, 80 b. & 200 mA& 160 turns & $9.6\times10^3$ turns \\
\end{tabular}
\end{ruledtabular}
\end{table}

\subsubsection{Beam Lifetime}
\begin{figure}
\includegraphics[width=1\linewidth]{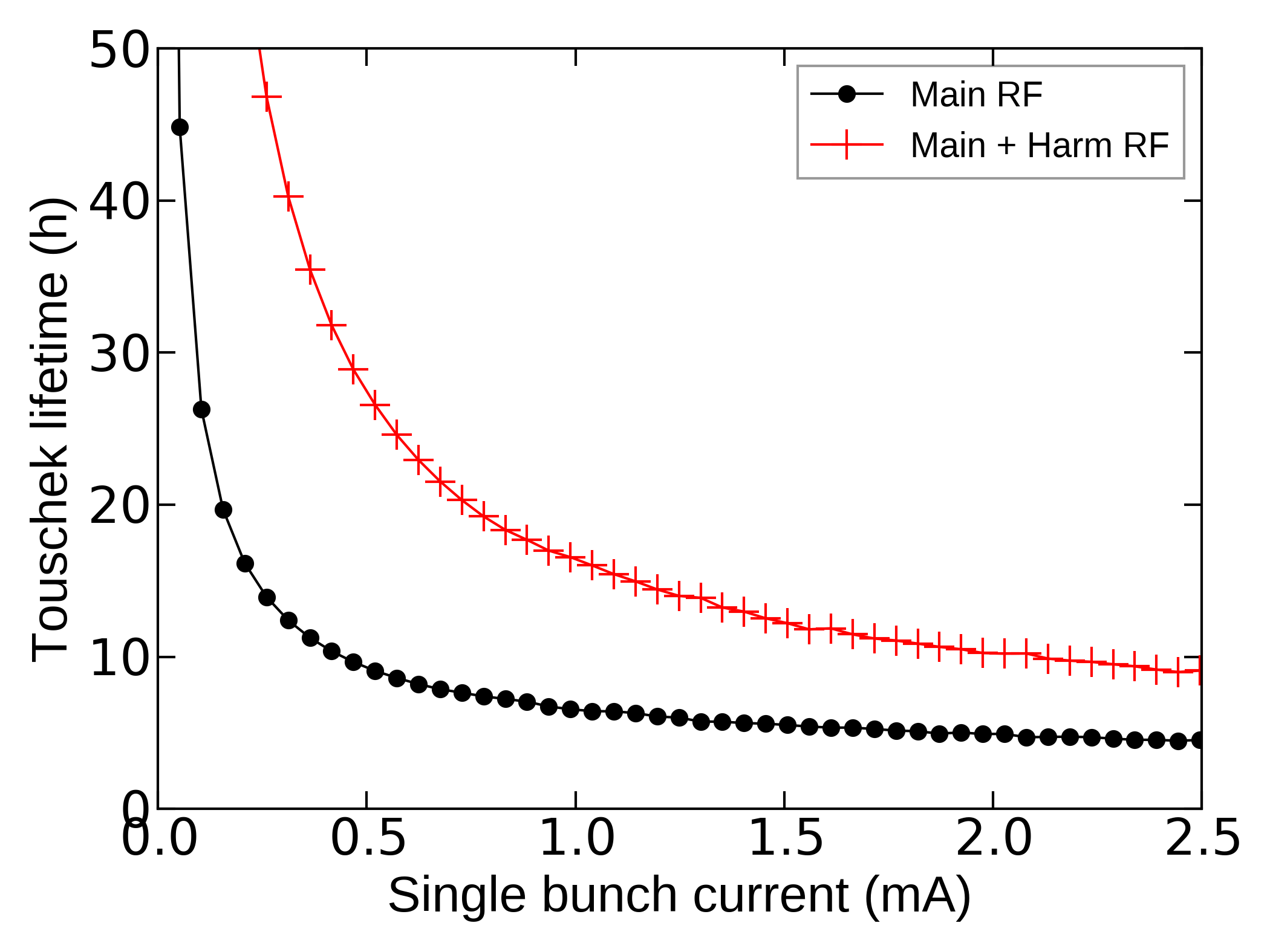}
\caption{\label{fig_touschlife} Touschek lifetime as a function of the single bunch current. The coupling is 10\%.}
\end{figure}

In an electron storage ring, the total beam lifetime can be estimated as $1/\tau=1/\tau_{TS} +1/\tau_{E} +1/\tau_{Ine}$, where $\tau_{TS}$, $\tau_{E}$ and $\tau_{INE}$ are contributions from the Touschek Scattering, the elastic and the inelastic scattering respectively.  The inelastic lifetime  has a week dependency on the momentum acceptance and scales approximately as $\propto \ln \left( 1/\delta_{acc} -5/8 \right)^{-1}$, and the elastic lifetime is mostly determined by the vertical machine acceptance $A_y$ and the average vertical function $\bar{\beta}_y$, and scales as  $ \propto A_y^2/ \bar{\beta}_y^2$ \cite{Moller:1999sp}. The Touschek lifetime $\tau_{TS}$   scales proportionally to the third power of the momentum acceptance and inversely proportionally to the bunch current $\propto \delta_{acc}^3/I_B$. Momentum acceptance optimization is thus important to guarantee good Touschek lifetime. In PETRA~IV, the Touschek lifetime as a function of the single bunch current is shown in Fig.~\ref{fig_touschlife}. The calculations were performed by the touschekLifetime toolkit \cite{xiao2007touschek}. Vacuum system design has the goal to achieve combined vacuum lifetime in access of 50 hrs. With this, the beam lifetime would be dominated by the vacuum scattering in the brightness mode with 0.1 mA bunch current, and by the Touschek scattering in the timing mode with 1 mA bunch current.

\subsection{Electron beam parameters}
Beam parameters including the effects of IBS, impedance, third-harmonic cavity and IDs are presented in Table~\ref{tab:beam_parameters} and  in Fig.~\ref{fig:SingleBunchLength}. Here 0.1~mA bunch current corresponds to the brightness and 1~mA per bunch to the timing mode.

\begin{figure}
\centering
\includegraphics[width=1\linewidth]{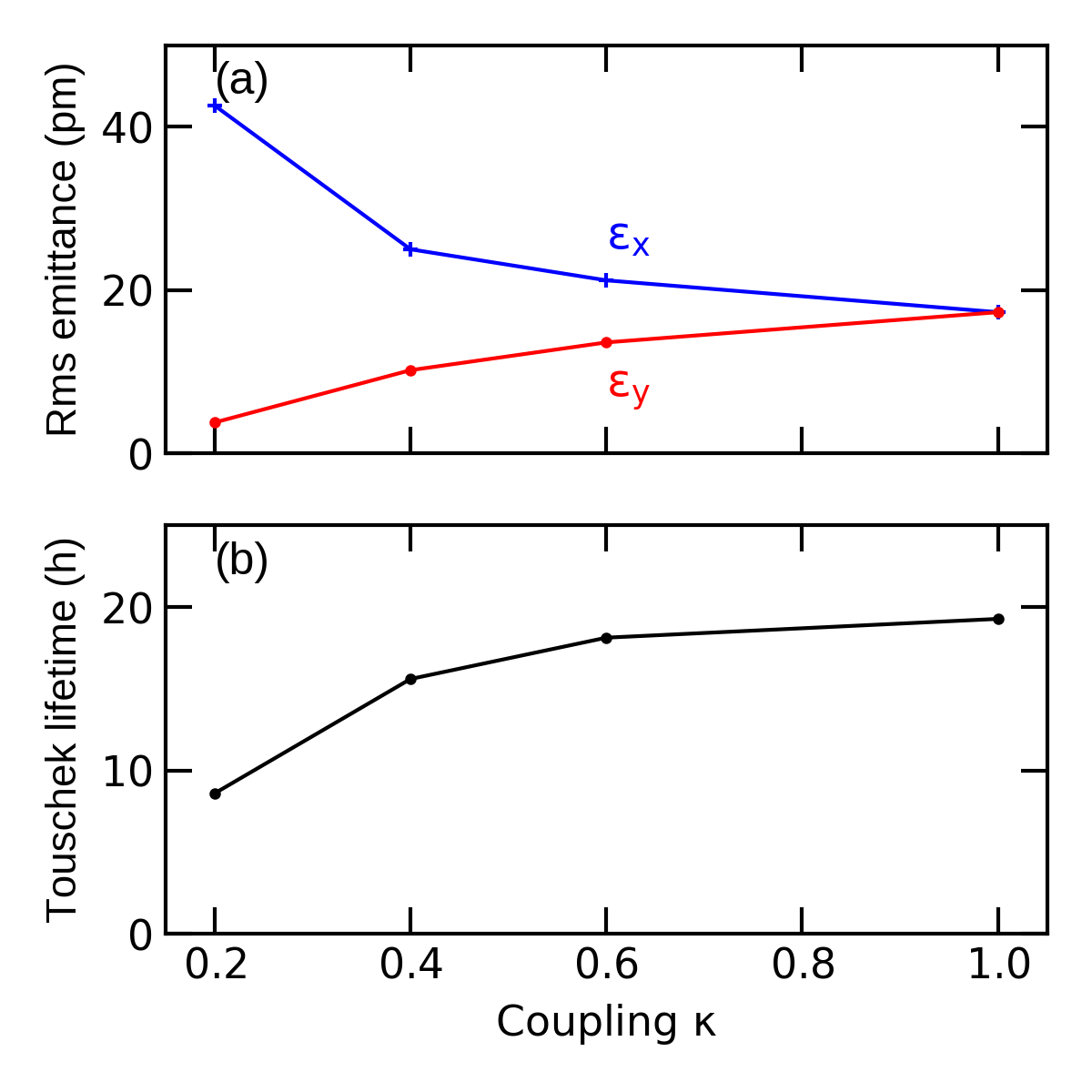}
\caption{\label{fig:coupling} Rms. emittance (a) and Touschek lifetime (b) as functions of coupling. Values are averages over statistical ensembles of lattices with up to 10\% beta beating.  Main and harmonic RFs are taken into account. The single bunch current is 1 mA.} 
\end{figure}

Nominal coupling of 10-20 \% is assumed for the baseline parameter evaluation. Larger coupling ratios are beneficial for improving the beam lifetime and mitigating the effect of intra-beam scattering.  Figure~\ref{fig:coupling} presents beam parameters for larger coupling ratios. The coupling could be achieved by operation close to the coupling resonance without impacting the beam dynamics performance as shown in \cite{Assmann:2023cdg}. The round or nearly round beam operation has so far not been found preferable by the experimental PETRA IV community due to the expected performance limitations of the X-ray optics. This capability could however be exploited in the future.

\begin{table}
\caption{\label{tab:beam_parameters} Electron beam parameters for different currents with coupling  $\kappa=10\%$.}
\begin{ruledtabular}
\begin{tabular}{lrrrr}
Bunch cur. (mA) & $\varepsilon_x$ (pm) & $\varepsilon_y$ (pm) & $\sigma_s$ (mm) & $\sigma_{\delta}$ ($10^{-3}$)  \\
\hline
0.01 & 18 & 1.8 & 11 & 0.9 \\
0.1 (Brightness) & 24 & 2.4 & 14 & 0.9 \\
1.0 (Timing) & 37 & 3.7 & 21 & 1.15 \\
\end{tabular}
\end{ruledtabular}
\end{table}

\subsection{Brilliance}

To increase the brilliance of the undulator radiation the beta function at the center of the ID straight should be close to the optimum value. For the beta functions at the ID center $\beta^\star = 2.2$~m was chosen in both planes as a compromise between small $\beta^\star$ and technically feasible gradient strength of the triplet magnets. 

\begin{figure}
\includegraphics[width=\linewidth]{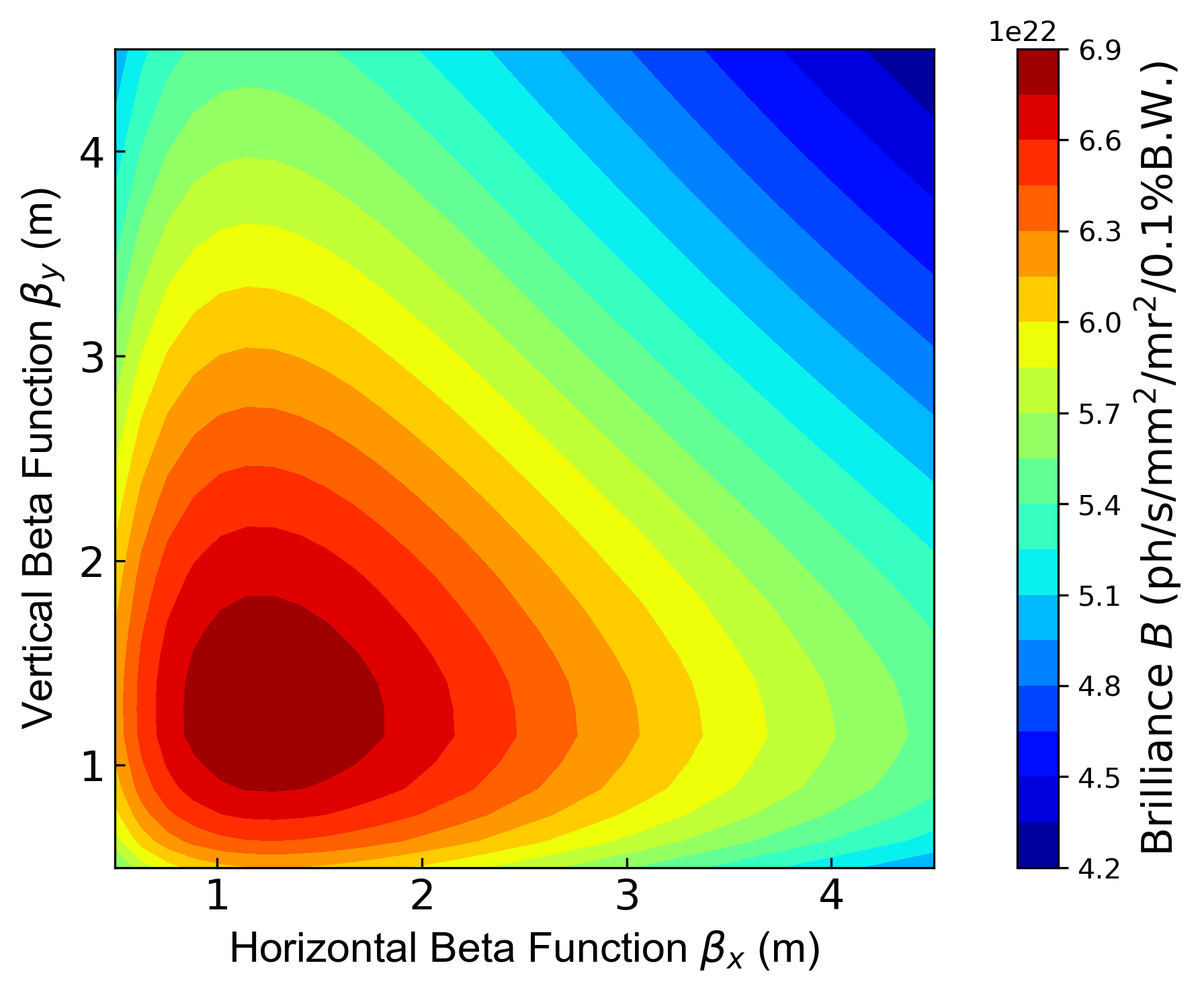}
\caption{\label{fig:brilliance-beta} Peak brilliance as a function of the beta functions at the center of the ID (U23 undulator, ID length 4.5~m, photon energy 10~keV, brightness mode).}
\end{figure}

The dependency of the peak brilliance on the beta function at the ID is shown in Fig.~\ref{fig:brilliance-beta} for a photon energy of 10~keV. The lattice parameters of the brightness mode with 20\,\% emittance ratio are used. An undulator with a period length of $23$~mm and an ID length of $L_{\textrm{ID}}= 4.5$~m is assumed. The calculation was done with SPECTRA~\cite{Tanaka:JSR_SPECTRA_2021} using the Wigner function approach. The optimum beta function would be $\beta \approx 1.1$~m which is smaller compared to  $\beta_{\textrm{opt}} = L_{\textrm{ID}}/\pi \approx 1.4$~m at 10~keV photon energy which is an approximation for a beam without energy spread~\cite{Walker:PRAB.22.050704}.

\begin{figure}
\includegraphics[width=\linewidth]{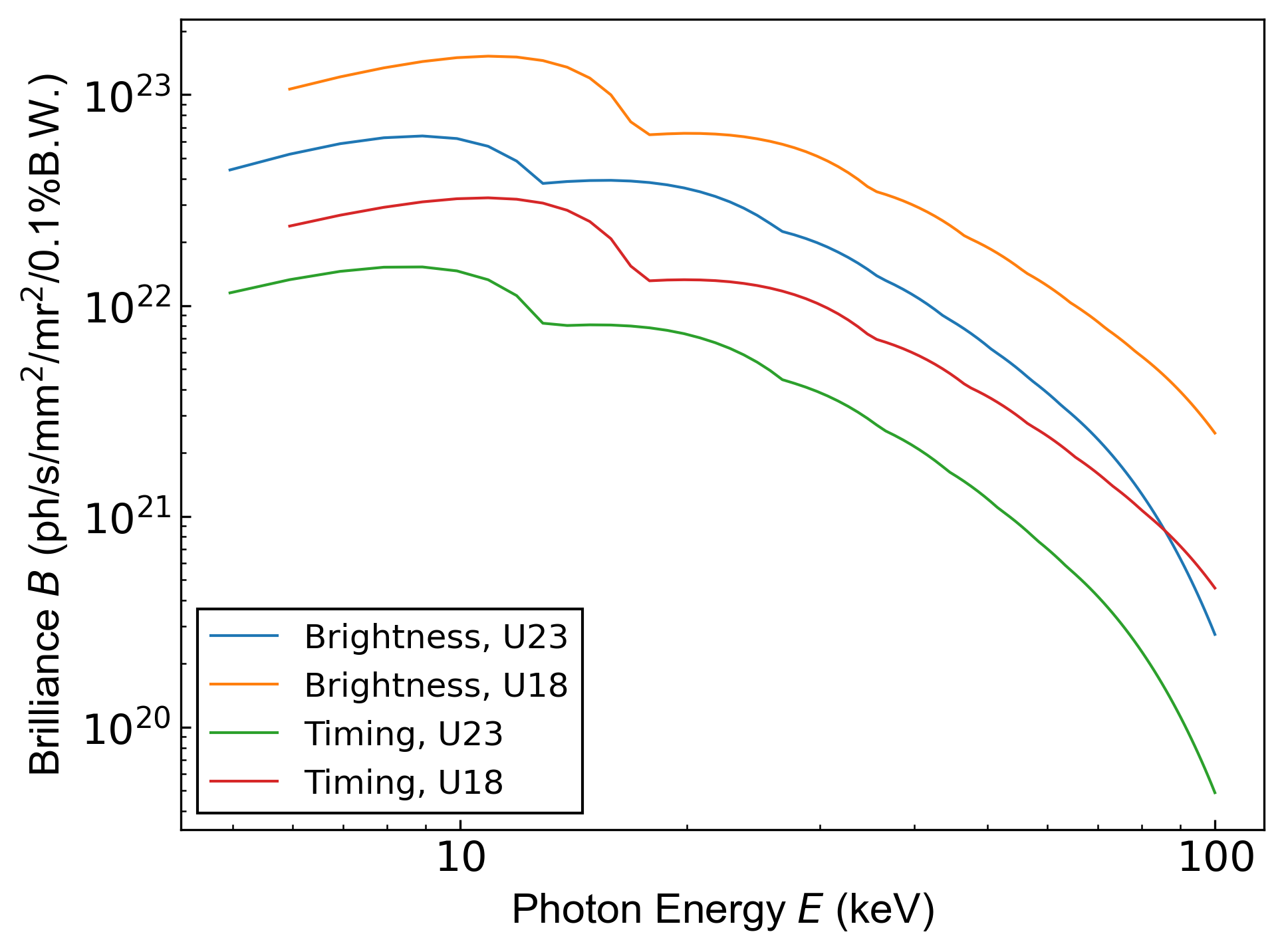}
\caption{\label{fig:brilliance-energy} Brilliance as a function of the photon energy for a U23 (4.5~m) and U18 (10~m) undulator for the brightness and timing mode.}
\end{figure}

The brilliance curve of a U23 undulator with a length of 4.5~m and a U18 undulator with a length of 10~m is shown in Fig.~\ref{fig:brilliance-energy} for the brightness and timing mode of PETRA~IV. 

\section{Parameters of main subsystems}
\label{sec:parameters}

Designs of most technical systems follow the concepts presented in the CDR \cite{Schroer2019}. Here we review the parameters of main subsystems.

\subsection{Magnets}

The magnet parameters are listed in Table~\ref{tab:magnets} and are such that the most challenging magnet is the high-gradient quadrupole. All preliminary magnet designs have been accomplished, for more detail see \cite{Bartolini:2022kgw}, \cite{Assmann:2023wyr}, \cite{Assmann:2023skl}, \cite{Assmann:2023fza}.

\begin{table}[h]
\caption{\label{tab:magnets} Magnet parameters }
\begin{ruledtabular}
\begin{tabular}{lrr}
Magnet type & Max. strength & Half-aper.  \\
\hline
Combined (Dip., Quad.) & 0.3~T, 12~T/m & 12.5 mm   \\
Quadrupole,triplet & 115 T/m & 11 mm\\
Quadrupole,cell & 100 T/m & 12.5 mm\\
Sextupole & $5\times 10^3$ T/m$^2$ & 12.5 mm\\
Octupole & $10^5$ T/m$^3$ & 12.5 mm \\
Orbit corrector & 1 mrad & 12.5 mm \\
Fast orbit corrector & 30 $\mu$rad, 1~kHz & 12.5 mm\\
\end{tabular}
\end{ruledtabular}
\end{table}

\subsection{RF system}

As discussed previously, PETRA IV will have a double-frequency RF system to lengthen the bunch and mitigate the IBS and Touschek effects. The RF system will feature HOM-damped cavities  with the parameters listed in Table~\ref{tab:rf}. The voltage was chosen such that the momentum acceptance is not limited by the RF bucket height and optimal bunch lengthening is achieved.

\begin{table}[h]
\caption{\label{tab:rf} Parameters of PETRA~IV RF systems}
\begin{ruledtabular}
\begin{tabular}{lr}
Parameter & Value \\
\hline
Energy loss per turn & 4.18  MeV   \\
Main RF frequency & 500  MHz   \\
Harmonic number & 3840 \\
Main RF total voltage  & 8.0  MV   \\
RF bucket height without IDs &  7.1 \% \\
RF bucket height with IDs &  4.8 \% \\
Harmonic RF frequency & 1.5  GHz \\
Harmonic RF total voltage  & 2.3  MV   \\
\end{tabular}
\end{ruledtabular}
\end{table}

\subsection{Injection chain}\label{sec:injection_chain}
The injection schemes adopted at light sources are the more usual off-axis injection with accumulation and the swap-out on-axis injection adopted at some new projects such as the APS-U \cite{apsu}.

\begin{figure}
\includegraphics[width=\linewidth]{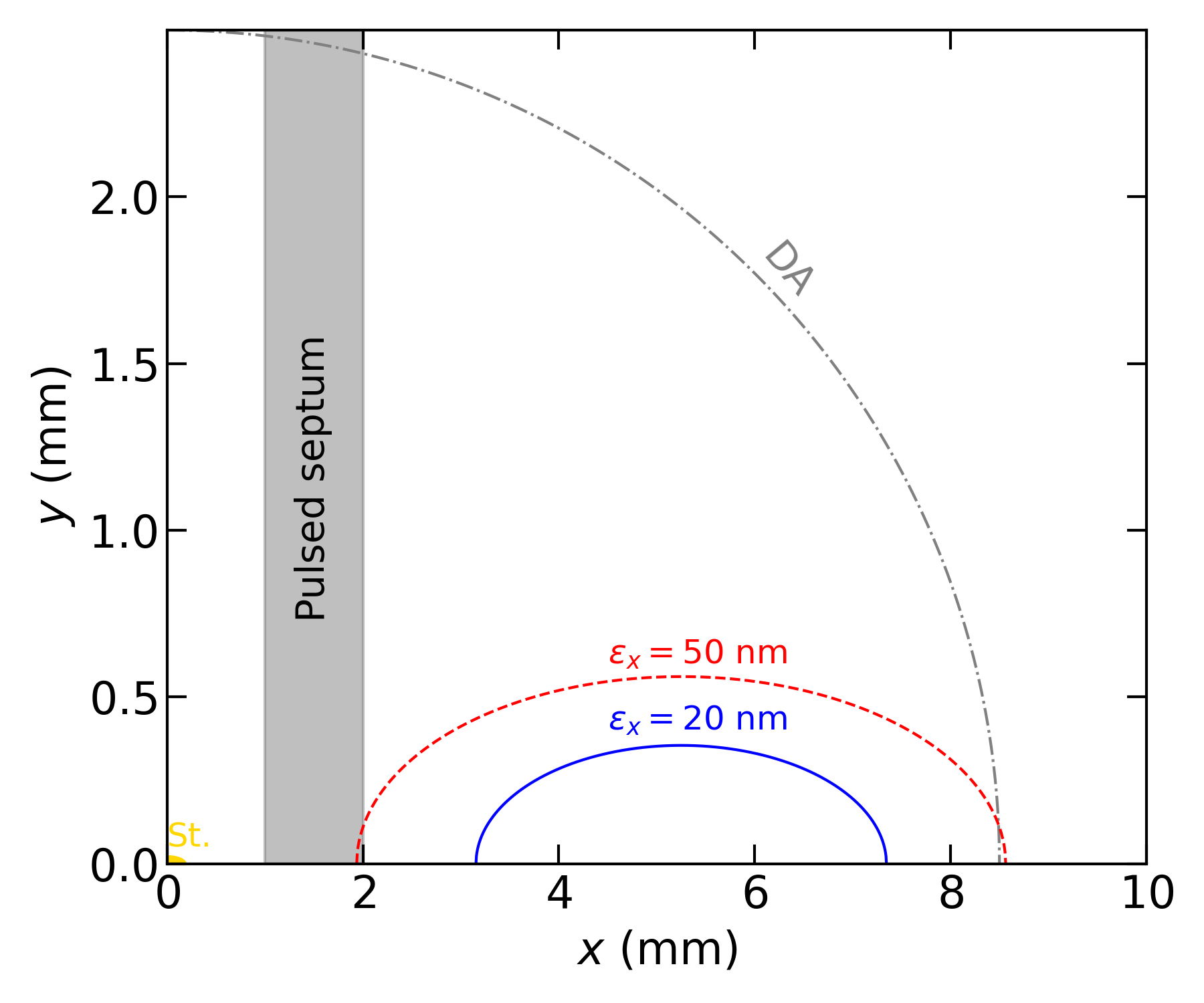}
\caption{\label{fig:beams_and_apertures}3 rms. beam sizes of the injected beam at the injection point during off-axis injection. The physical aperture of the pulsed septum and the dynamic aperture with realistic errors are also shown. The origin corresponds to the bumped orbit of the stored beam (``St.''), its 6 rms. beam size is shown in yellow.}
\end{figure}

\begin{figure}
\includegraphics[width=.99\linewidth]{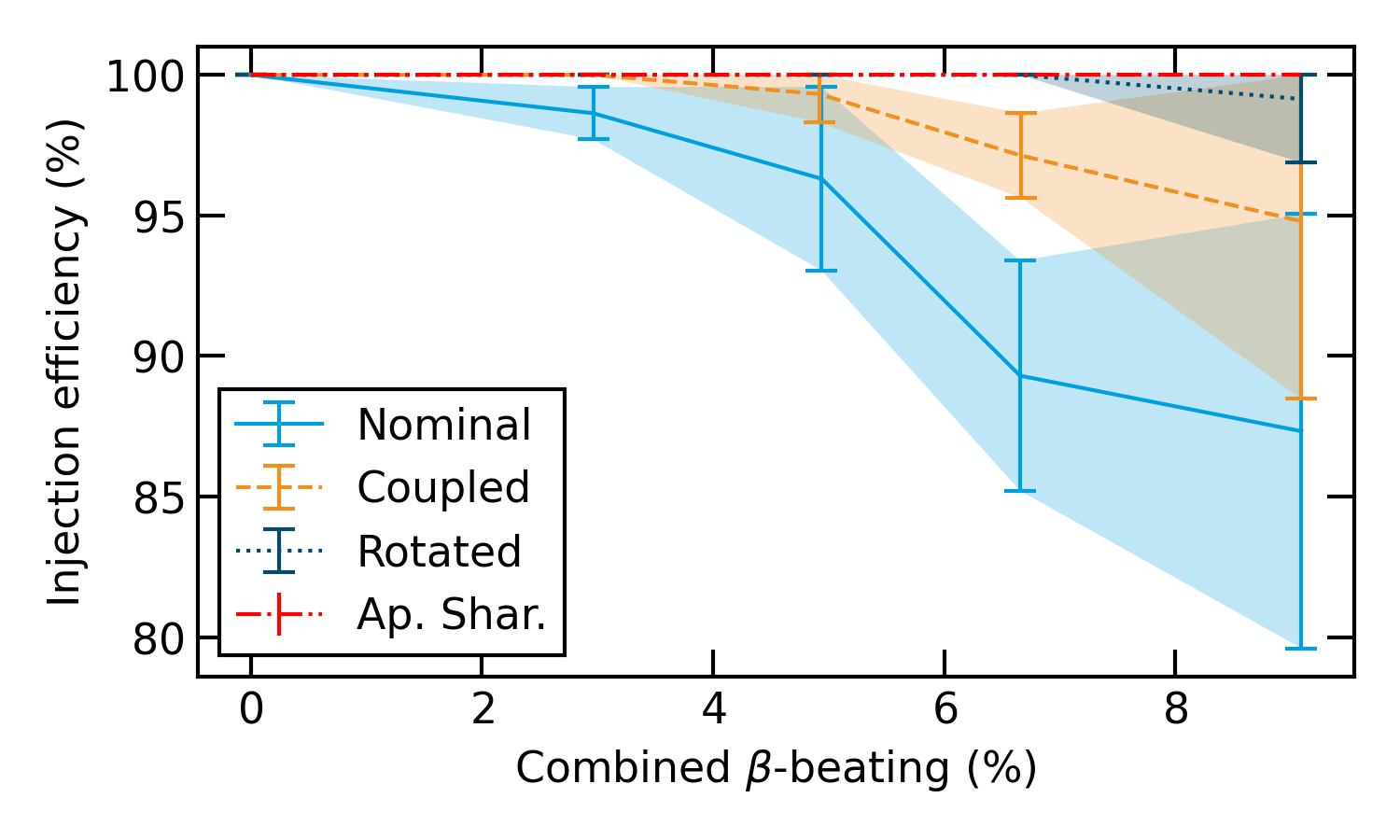}
\caption{\label{fig:injection_efficiency} Injection efficiency versus beta beating: lines correspond to average expected values and error bars and shaded areas depict the spread between error seeds. Nominal beam: emittances 19 nm rad/1.9 nm rad, offset 8 mm, $\beta_x$ 19.9 m; coupled beam: emittance 10.4 nm rad /10.4 nm rad, offset 7.5 mm, $\beta_x$ 17.6 m;
rotated beam: emittance 1.9 nm rad /19 nm rad, offset 6.6 mm, $\beta_x$ 11.5 m;
aprture sharing: emittance 19 nm rad /1.9 nm rad, offset 3.2 mm, $\beta_x$ 29.6 m.
}
\end{figure}

With the dynamic aperture (with errors) of $\pm$ 8 mm and $\beta_x$ = 46~m at the injection the maximal possible booster emittance for the on-axis injection is about 150~nm, assuming 3$\sigma$ acceptance. Further, assuming a 5~mm beam separation required for a thin pulsed septum, space for the stored beam, and orbit error margins the maximum possible booster emittance would be about 50 nm for the off-axis accumulation (Fig.~\ref{fig:beams_and_apertures}). With 20 nm emittance the new DESY IV booster should satisfy these requirements \cite{Chao:2021fyh, Chao:2021jru} with some margin. DESY IV parameters are shown in Table \ref{tab:booster}. To take into account more subtle effects of energy spread, bunch length and momentum acceptance, injection efficiency simulations were performed. Perturbed lattices with different magnitudes of alignment errors were generated and corrected, and bunches with various initial offsets and particle distributions corresponding to the DESY IV booster parameters were generated and tracked with the {\it Elegant } code \cite{elegant}. Following scenarios were studied: on-axis injection, off-axis injection with closed injection bump and off-axis injection with shared oscillations between the injected and the stored beams. Moreover, the Twiss parameters at the end of the transfer line were optimized for best phase-space matching of the injected beam to the acceptance; these parameters differ from the values of the ring optics at the septum, where $\beta_x$=46 m and $\alpha_x$=0.
The resulting injection efficiency as a function of the combined horizontal and vertical beta beating is presented in Figure \ref{fig:injection_efficiency}. Based on experience at other machines and on the results of correction simulations, we expect the machine to operate with 1-2 \% beta beating. Injection aperture sharing and beam rotation in the transfer line would allow for 100\% injection efficiency under much more pessimistic optics correction assumptions.

\begin{table}[ht]
\caption{\label{tab:booster} Booster parameters }
\begin{ruledtabular}
\begin{tabular}{lll}
Parameter & 6-fold option & 8-fold option \\
\hline
Circumference  & 316.8 m  & 304.8 m \\
Injection energy  & 450 MeV  & 450 MeV \\
Extraction energy & 6 GeV  & 6 GeV\\
Repetition rate & 5 Hz  & 5 Hz \\
Natural emittance  & 19 nm rad & 21 nm rad  \\
Bunch length at extraction & 20 mm & 17 mm\\
Energy loss/turn & 6.55 MeV & 6.67 MeV\\
RF voltage & 12 MV & 12 MV\\
RF frequency & 500 MHz & 500 MHz\\
Rel. energy spread  & 0.27 \%  & 0.22 \% \\
Single bunch charge  & up to 1 nC & up to 1 nC   \\
\end{tabular}
\end{ruledtabular}
\end{table}

On-axis swap-out injection would require the injection chain to accelerate and deliver charges of approx. 20 nC. While this has been shown to be possible in simulations \cite{Chao:2021fyh}, a low charge operation is considered more safe and is chosen as baseline. Injection aperture sharing, which has larger perturbation on science experiments, could be adopted in case the standard off-axis injection has insufficient efficiency due to additional errors not taken into account in present simulations. 

\begin{figure}
\includegraphics[width=.99\linewidth]{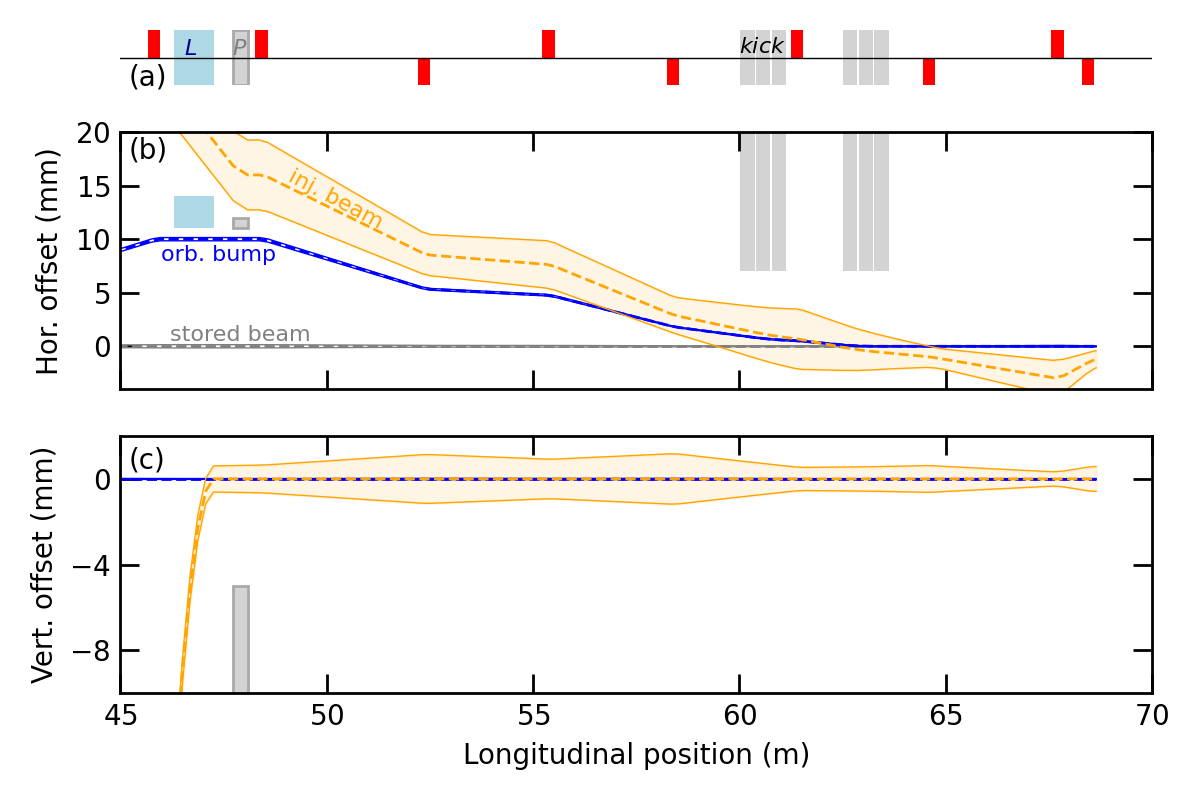}
\caption{\label{fig:injection_schematic} 
Schematic of the injection region (a) and beam trajectories (b,c). Quadrupole magnets are shown in red, the Lambertson septum ($L$) in light-blue, the thin pulsed septum ($P$) and injection stripline kickers ($kick$) in grey. Trajectories of the stored beam (grey), bumped beam (blue), and injected beam (orange) are plotted in dashed lines together with 5 rms. beam sizes, shown as shaded areas.
Aperture restrictions created by the injection devices, kickers and septa, are shown in their respective color.
}
\end{figure}

The injection chain would also comprise an S-Band linac with the energy of 450 MeV, an accumulator ring mostly used to damp the energy spread of linac bunches, and the booster synchrotron. 

\section{Possible advanced capabilities}
\label{sec:advanced}

The distinctive feature of PETRA IV is the presence of very long straight sections that correspond to approximately 30\% of the machine circumference, and about 40\% of the arc space being unavailable for installation of insertion devices due to civil engineering constraints. 
This is unusual for dedicated synchrotron radiation facilities. Smaller colliders  such as DORIS \cite{Nesemann:1995jd} and SPEAR \cite{Cantwell:1993js} have been converted to radiation sources in the 1990s after exhausting their high-energy physics potentials, and several proposals for refurbishing higher energy colliders have been discussed \cite{Cai:PhysRevSTAB.15.054002, Borland:2012zz}. However, such conversion is associated with a number of limitations, and the majority of modern synchrotron radiation sources have been constructed as green field facilities. 

The distinct geometry has been exploited for optimizing machine performance, as has been discussed in more detail previously: the short straight sections where user insertion devices could not be placed are used to optimally install damping wigglers and provide more than a factor of two emittance reduction; a long straight section provides sufficient place for compact first- and third-harmonic RF system installation; injection section could be optimized for large beta function that minimizes the impact of the septum blade thickness on the injection efficiency; ample space is available to install the multi-bunch feedback systems and fast injection stripline kickers.   

The geometry could also be used for unique radiation generation capabilities. So, the photon flux scales linearly with the undulator length and the very long straight sections could be used to install very long (up to 80 m) undulators. PETRA III already features an approximately 10 m long device \cite{Wille2010}, and SPring-8, another facility whose lattice provides space for very long devices, 
an undulator of approximately 23 m length \cite{YABASHI2001678}. Such extremely long devices have several disadvantages. First, the minimum gap value is limited by the stay-clear that scales linearly with the undulator length and inversely with the square root of the beta function at the center of the undulator $\sim L_{und}/\sqrt{\beta^{\star}}$. As the gain in undulator field is exponential with the gap reduction \cite{Halbach:1982vx}, reducing the gap turns out to be more beneficial than increasing the undulator length. A possible approach to providing a well-focused electron beam in a long undulator that is compatible with a small gap size is to interleave the undulator sections with focusing magnets, the approach usual at free-electron lasers. This, unfortunately, leads to multiple radiation source points and loss in brightness. Optimization of the undulator parameters depends on the application and is outside of the scope of this article. PETRA IV provides five slots for the  so-called flagship devices, that could be of the maximum length of 10 m.

The combination of very low emittance with the possibility to install a very long undulator suggests the possibility of using the ring to drive an x-ray free-electron laser. Indeed, this possibility has been explored, with the outcome that in the high gain regime only relatively soft x-ray wavelengths would be accessible, only in the exponential regime with large intensity fluctuations, and only after installing significant additional RF for longitudinal bunch compression \cite{Agapov:2015dha}. Further, the possibility of the ring to drive an x-ray oscillator (XFELO) has been studied \cite{Agapov:2018cse}. It has been shown that the electron beam quality is sufficient for x-ray cavity pumping and for build-up of the lasing. However, the scheme relied on installing the XFELO in an electron beam by-pass where the bunches would be kicked with a certain frequency high enough to pump the cavity but low enough to let the energy spread and emittance growth induced by the beam-cavity interaction damp.
This is technically possible but not practically feasible due to a large amount of additional installations and interference with the operation of other beamlines. Recently, the scheme was further developed to show that the operation is possible without the by-pass, but through charge stacking of individual bunches, i.e. adjusting the ring fill pattern to have a few high-charge bunches that would be pumping the cavity with approximately 2 MHz frequency \cite{Li:2023vfu}. Further feasibility studies of such an option for the PETRA IV ring are required. 

Finally, recent studies have shown that PETRA IV can be a source of powerful coherent THz radiation  \cite{Antipov:2023rjd}. A corrugated metallic or a dielectric structure could be used to induce self-wakes that produce stable bunch shortening and produce radiation that can be outcoupled for experimental purposes.

\begin{acknowledgments}
We wish to acknowledge contributions of P.~Raimondi and S.~Liuzzo to establishing the lattice concept.
\end{acknowledgments}
\newpage 

\bibliography{combibib}

\end{document}